\documentclass[12pt]{article}
\usepackage{amssymb}
\usepackage{amsmath}
\usepackage{graphicx}
\usepackage{indentfirst}
\usepackage{cite}

\begin{document}

\title{Cross sections for meson-meson nonresonant reactions}
\author{Yu-Qi Li \and Xiao-Ming Xu}
\date{}
\maketitle \vspace{-1cm} 
\centerline{Department of Physics, Shanghai
University, Baoshan, Shanghai 200444, China}
%----------------------------------------------------------------
\begin{abstract}
Meson-meson nonresonant reactions governed by the quark-interchange mechanism
are studied in a potential that is derived from QCD. S-wave elastic phase 
shifts for $I=2~\pi\pi$ and $I=3/2~K \pi$ scattering are obtained with wave
functions determined by the central spin-independent term of
the potential. The reactions include inelastic 
scatterings of two mesons in the ground-state pseudoscalar octet and the
ground-state vector nonet. Cross sections for reactions involving $\pi$, 
$\rho$, $K$ and $K^\ast$ indicate that mesonic interactions in matter 
consisting of only $K$ and $K^\ast$ can be stronger than mesonic interactions
in matter consisting of only $\pi$ and $\rho$ and the reaction of 
$I=3/2$ $\pi K^\ast \to \rho K$ is most important among the 
endothermic nonresonant reactions. By the quark-interchange mechanism
we can offer $\sqrt s$-dependences of $\phi$ absorption cross sections in 
collisions with $\pi$ and $\rho$ and relevant average cross sections what are
very small for the reaction of $I=1$ $\pi \phi \to K^{\ast} K^{\ast}$ and
remarkably large for the reaction of $I=1$ $\rho \phi \to K^{\ast} K^{\ast}$.
It is found from the $\sqrt s$-dependences of cross sections that $\rho$ and
$K^{\ast}$ creation cross sections can be larger than their absorption cross
sections, respectively.
\end{abstract}



\newpage

\leftline{\bf 1. Introduction}
\vspace{0.5cm}
Ratios of $p_T$-integrated mid-rapidity yields for mesons measured in PHENIX
\cite{adler} for central Au+Au collisions at $\sqrt {s_{NN}}=200$ GeV are
$\pi^-/\pi^+ =0.984$, $K^-/K^+ =0.933$, $K^+/\pi^+ =0.171$ and $K^-/\pi^- 
=0.162$, which agree with the measurements of other collaborations   
\cite{arsene,back,adams1,adcox}. The ratio $\rho^0/\pi^- =0.169$ is obtained
by the STAR Collaboration \cite{adams2} in peripheral Au+Au collisions at
$\sqrt {s_{NN}}=200$ GeV. The ratios $\pi^-/\pi^+$ and $K^-/K^+$ keep
unchanged at $p_T \leq 2$ GeV/$c$ or over the whole centrality range 
but $K^+/\pi^+$ and
$K^-/\pi^-$ increase with $p_T$ or centrality \cite{adler}. The ratios  
$K^-/\pi^-$ and $K^-/K^+$ do not vary near midrapidity and 
decrease slowly in the other measured rapidity region while both
$\pi^-/\pi^+$ and $K^+/\pi^+$ are constants at $-3<y<3$ \cite{arsene,bearden}.
It is evident from the above experimental results that pions, kaons, rhos are
dominant meson species in hadronic matter. Subsequently, 
extended attention must be paid from the reactions $\pi \pi \to \pi \pi$ and 
$\pi \pi \to K \bar K$ in hadronic physics to
a great number of other meson-meson reactions occurred in hadronic matter. 
On one hand meson-meson cross sections are crucial to chemical equilibration,
thermalization, hadron flows and hadron yields, on the other hand
earlier decoupled mesons due to small cross sections can show relatively
clear information on quark-gluon plasma. Therefore, the study of
meson-meson scatterings is a fundamental task in RHIC heavy-ion collisions.

A typical study of elastic $\pi \pi$ scattering is offered in chiral 
perturbation theory since pions represent Goldstone bosons.
Phase shift for the elastic scattering at low energy can be well
accounted for up to two loops that receive contributions from ${\cal L}_2$,
${\cal L}_4$ and ${\cal L}_6$ of the chiral Lagrangian for pion fields
while the $\pi \pi$
scattering amplitude is constrained by analyticity, unitarity, crossing
symmetry and chiral symmetry \cite{Gasser}. 
A combined study of the $I=0$ $\pi \pi \to \pi \pi$ and $\pi \pi \to K \bar K$
with Lippmann-Schwinger equations in coupled channels 
could also reproduce the phase shifts and inelasticities 
in the lowest order chiral Lagrangian that includes the pseudoscalar octet
\cite{OO}.

In order to explain the measured elastic phase shifts 
\cite{Col71,Dur73,Hoo77,Los74,Jon73,Est78} for $I=2~\pi \pi $ and 
$I=3/2~K \pi$ scattering, 
Barnes {\it et al.} \cite{BS92,BSW92} considered Born-order 
quark-interchange diagrams in a nonrelativistic potential model and few other 
attempts have been made \cite{Wei91,Zhao98}. Motivated by the important 
$J/\psi$ 
suppression problem, the quark-interchange model in its reasonable form
has been applied to study hadron-$J/\psi$ dissociation cross sections 
\cite{WSB,BSWX}. To the current need of obtaining unknown cross sections for
meson-meson reactions in hadronic matter, the quark-interchange mechanism with 
a potential derived from QCD will at the present work produce new and useful 
results that include cross sections for the nonresonant reactions of
 $I=2$ $\pi \pi \leftrightarrow 
\rho \rho$, $I=2$ $\pi \rho \leftrightarrow \rho \rho$,    
 $I=1$ $KK \leftrightarrow K^\ast K^\ast$,
 $I=1$ $KK^\ast \leftrightarrow K^\ast K^\ast$,
 $I=\frac {3}{2}$ $\pi K \leftrightarrow \rho K^\ast$,
 $I=\frac {3}{2}$ $\pi K^\ast \leftrightarrow \rho K^\ast$,
 $I=\frac {3}{2}$ $\rho K \leftrightarrow \rho K^\ast$ and
 $I=\frac {3}{2}$ $\pi K^\ast \leftrightarrow \rho K$.
These cross sections are unknown in both experiment and theory. The reactions
originate from the quark-interchange process.

The $\phi$ mesons relevant to strangeness enhancement are used to conclude
that the particle-identified elliptic flow coefficients have the scaling
behavior over the constituent-quark number \cite{Ma,Chen,CK}. Results of the
$\phi$ meson measurements \cite{phi} 
are usually explained with the assumption that hadron-$\phi$ cross
sections are so small that $\phi$ mesons will retain information from the 
early hot and dense matter \cite{Sho85,jjd,Bra03}. 
Accordingly, it is very interesting to examine the assumption in theory. 
Meson-phi interactions may lead to collisional broadening of the phi meson 
width in spite of the small meson-phi cross sections that were obtained from   
effective meson Lagrangians \cite{BR,KS,Haglin}. The Lagrangians also give 
$\phi$ mean free path larger than that obtained from large collision rates of 
phi meson with $\rho$, $K$ and $K^*$ mesons as a result of
the hidden local symmetry 
Lagrangian \cite{AK}. The discrepancy created from the two Lagrangians is 
due to the different evaluations of hadron-phi scatterings. 
To help understand the behavior of $\phi$ meson in hadronic matter, the 
hadron-phi cross sections need to be studied in different ways, especially in 
the attempt of quark level, which is just the second purpose of the present 
work. We calculate cross sections for the nonresonant reactions of
 $I=1~\pi \phi \to K K^\ast$ (or $K^\ast K$),
 $I=1~\pi \phi \to K^\ast K^\ast$,
 $I=1~\rho \phi \to K K$, $I=1~\rho \phi \to K
K^\ast$ (or $K^\ast K$) and  $I=1~\rho \phi \to K^\ast K^\ast$.
These cross sections have not been evaluated in quark models. The reactions 
arise out of the quark-interchange process.

In next section we describe the potential derived from QCD and mesonic 
quark-antiquark wave functions. Formulas established with the
quark-interchange mechanism are presented in Section 3. Elastic phase shifts 
for $I=2$ $\pi\pi$ and $I=3/2$ $K\pi$ scattering, cross sections for the 
nonresonant reactions and discussions are given in Section 4. 
Conclusions are in the last section.

\vspace{0.5cm}
\leftline{\bf 2. Potential and radial wave functions}
\vspace{0.5cm}

The Buchm\"{u}ller-Tye potential \cite{Buc81} is a QCD-like potential 
because the running coupling constant is obtained from the $\beta$ function
that takes a linear form in the strong-coupling limit is 
renormalization-scheme-independent and gauge-invariant up to two loops in the 
weak-coupling region and contains a term related to the three-loop 
contribution.
The potential was nonrelativistic, central and spin-independent. However, the 
running coupling constant leads to a relativistic potential that takes as the 
first term the linear confinement and as the second term a relativistic
one-gluon-exchange potential with the perturbative one- and two- loop 
corrections. In the second term the part responsible for the confinement is 
subtracted from the running coupling constant. We peform nine times of 
different Foldy-Wouthuysen canonical transformations to the two-constituent
Hamiltonian with the relativistic potential. A lengthy calculation gives a 
central spin-independent term, spin-spin interaction, 
spin-orbit and tensor terms \cite{Xu02}. In this way the 
Buchm\"{u}ller-Tye potential is extended. The central spin-independent term, 
i.e. the Buchm\"{u}ller-Tye potential, contains the linear confinement and loop
corrections to the Coulomb potential. The term does not rely on the masses of 
quark and antiquark constituents. The spin-spin interaction includes a 
contact term and a loop-correction term. The spin-orbit and tensor terms 
contain factors that are given by loop corrections. The spin-spin, spin-orbit
and tensor terms are related to the inverse of constituent mass.
The spin-orbit and tensor terms do not contribute to the masses of mesons  
in the ground-state pseudoscalar octet and the ground-state vector nonet. 
The central spin-independent term is
\begin{equation}
V_{\rm si}= - \frac {\vec {\lambda}_a}{2} \cdot \frac {\vec
{\lambda}_b}{2} \frac {3}{4} {\rm k}r+\frac {\vec {\lambda}_a}{2}
\cdot \frac {\vec {\lambda}_b}{2} \frac {6\pi}{25} \frac {v(\lambda r)}{r}
\end{equation}
where $\vec {\lambda}_a$ and $\vec {\lambda}_b$ are the Gell-Mann 
"$\lambda$-matrices", ${\rm k}=1/2\pi \alpha'$
and $\lambda=\sqrt {3b_0/16\pi^2\alpha'}$ with the Regge slope
$\alpha'=1.04{\rm GeV}^{-2}$. The function $v$ is 
\begin{equation}
v(x)=\frac
{4b_0}{\pi} \int^\infty_0 \frac {dQ}{Q} (\rho (\vec {Q}^2) -\frac {K}{\vec
{Q}^2}) \sin (\frac {Q}{\lambda}x)
\end{equation}
where $b_0=11-\frac {2}{3} N_f$ with the quark flavor number $N_f=4$,
$K= 3/16\pi^2\alpha'$ and $\rho (\vec {Q}^2)$ with gluon momentum $\vec Q$
is the physical running 
coupling constant given by Buchm$\rm \ddot u$ller and Tye \cite{Buc81}. 
The Schr$\rm \ddot o$dinger equation with the central spin-independent
potential produces a radial wave function for the quark-antiquark relative 
motion of the $\pi$ and $\rho$ mesons 
while both the up-quark and down-quark masses take the value 0.33 GeV. The
wave function leads to 0.6308 GeV for the mass splitting between $\pi$ and 
$\rho$, versus the experimental value 0.6304 GeV, that is calculated 
with the spin-spin interaction
\begin{equation}
V_{\rm ss}=
- \frac {\vec {\lambda}_a}{2} \cdot \frac {\vec {\lambda}_b}{2}
\frac {16\pi^2}{25} \delta^3(\vec {r})  \frac {\vec {s}_a \cdot \vec
{s}_b}{m_am_b}
+ \frac {\vec {\lambda}_a}{2} \cdot \frac {\vec {\lambda}_b}{2}
  \frac {4\pi}{25} \frac {1}{r}
\frac {d^2v(\lambda r)}{dr^2} \frac {\vec {s}_a \cdot \vec {s}_b}{m_am_b}
\end{equation}
where $\vec {s}_a$ ($\vec {s}_b$) and $m_a(m_b)$ are the spin and mass of
the quark or antiquark constituent $a(b)$, 
respectively. If the strange quark mass is 0.53 GeV 
and the radial quark-antiquark relative-motion wave functions of
$K$, $K^\ast$, $\eta$, $\omega$ and $\phi$ are the same as that of $\pi$ and 
$\rho$, we get the mass splittings $m_{K^\ast}-m_K=0.3928~{\rm GeV}$ and 
$\frac {1}{3}m_{\omega}+\frac {2}{3}m_{\phi}-m_\eta=0.3733~{\rm GeV}$ in
comparison to the experimental values 0.3963 GeV and 0.3930 GeV, respectively.
Here $m_K$, $m_{K^\ast}$, $m_{\eta}$, $m_{\omega}$ and $m_{\phi}$ are the
masses of $K$, $K^\ast$, $\eta$, $\omega$ and $\phi$, respectively. 
The reasonable agreement between the theoretical results and the experimental 
data allows us to assume that all the mesons in the ground-state pseudoscalar 
octet and the ground-state vector nonet have the same spatial
wave function of the quark-antiquark relative motion.

\vspace{0.5cm}
\leftline{\bf 3. Formulas}
\vspace{0.5cm}

In view of the fact that a meson consists of quark and antiquark constituents,
$A(q_{1}\overline{q}_{1})+B(q_{2}\overline{q}_{2})\to
C(q_{1}\overline{q}_{2})+D(q_{2}\overline{q}_{1})$ is a 4-to-4 constituent
scattering that cannot be easily treated in QCD. However, in nonrelativistic
dynamics what separates the quark-antiquark motion into their center-of-mass 
motion and relative motion, the scattering is reduced to a 2-to-2 
scattering of two quark-antiquark pairs and the scattering problem is
solved in a clear way from the S-matrix.
S-matrix element for the reaction 
$A(q_{1}\overline{q}_{1})+B(q_{2}\overline{q}_{2})\to
C(q_{1}\overline{q}_{2})+D(q_{2}\overline{q}_{1})$ is
\begin{equation}
S_{\rm fi} = \delta_{\rm fi} - 2\pi i \delta (E_{\rm f} - E_{\rm i})
<q_1\bar {q}_2,q_2\bar {q}_1 \mid H_{\rm I} 
\mid q_1\bar {q}_1,q_2\bar {q}_2> 
\end{equation}
where $H_{\rm I}$ is the interaction of two constituents each in an initial 
meson or in a final meson and $E_{\rm i}$ ($E_{\rm f}$) is the 
total energy of the two initial (final) mesons. 
The wave function of the initial mesons is 
\begin{equation}
\psi_{q_1\bar {q}_1, q_2\bar {q}_2}=
\frac {e^{i\vec {P}_{q_1\bar {q}_1}\cdot \vec {R}_{q_1\bar {q}_1}}}{\sqrt V} 
\psi_{q_1\bar {q}_1} (\vec {r}_{q_1\bar {q}_1})
\frac {e^{i\vec {P}_{q_2\bar {q}_2}\cdot \vec {R}_{q_2\bar {q}_2}}}{\sqrt V} 
\psi_{q_2\bar {q}_2} (\vec {r}_{q_2\bar {q}_2})
\end{equation}
which is normalized to one in volume $V$, 
\begin{equation}
\int d^3 r_{q_1\bar {q}_1} \psi^+_{q_1\bar {q}_1} (\vec {r}_{q_1\bar {q}_1})
\psi_{q_1\bar {q}_1} (\vec {r}_{q_1\bar {q}_1}) =1
\end{equation}
and
\begin{equation}
\int d^3r_{q_2\bar{q}_2}\psi^+_{q_2\bar{q}_2}(\vec{r}_{q_2\bar{q}_2})
\psi_{q_2\bar{q}_2}(\vec{r}_{q_2\bar{q}_2})=1.
\end{equation}
$\vec {P}_{q_1\bar {q}_1}$ ($\vec {P}_{q_2\bar {q}_2}$),
$\vec {R}_{q_1\bar {q}_1}$ ($\vec {R}_{q_2\bar {q}_2}$) and
$\vec {r}_{q_1\bar {q}_1}$ ($\vec {r}_{q_2\bar {q}_2}$) are the total 
momentum, the center-of-mass coordinate and the relative coordinate of $q_1$
($q_2$) and $\bar {q}_1$ ($\bar {q}_2$), respectively.
The wave function of the final mesons is
\begin{equation}
\psi_{q_1\bar {q}_2, q_2\bar {q}_1}=
\frac {e^{i\vec {P}_{q_1\bar {q}_2}\cdot \vec {R}_{q_1\bar {q}_2}}}{\sqrt V} 
\psi_{q_1\bar {q}_2} (\vec {r}_{q_1\bar {q}_2})
\frac {e^{i\vec {P}_{q_2\bar {q}_1}\cdot \vec {R}_{q_2\bar {q}_1}}}{\sqrt V} 
\psi_{q_2\bar {q}_1} (\vec {r}_{q_2\bar {q}_1})
\end{equation}
which is normalized to one in volume $V$,  
\begin{equation}
\int d^3r_{q_1\bar {q}_2} \psi^+_{q_1\bar {q}_2} (\vec {r}_{q_1\bar {q}_2})
\psi_{q_1\bar {q}_2} (\vec {r}_{q_1\bar {q}_2}) =1
\end{equation}
and 
\begin{equation}
\int d^3r_{q_2\bar {q}_1} \psi^+_{q_2\bar {q}_1} (\vec {r}_{q_2\bar {q}_1})
\psi_{q_2\bar {q}_1} (\vec {r}_{q_2\bar {q}_1}) =1.
\end{equation}
$\vec {P}_{q_1\bar {q}_2}$ ($\vec {P}_{q_2\bar {q}_1}$),
$\vec {R}_{q_1\bar {q}_2}$ ($\vec {R}_{q_2\bar {q}_1}$) and
$\vec {r}_{q_1\bar {q}_2}$ ($\vec {r}_{q_2\bar {q}_1}$) are the total 
momentum, the center-of-mass coordinate and the relative coordinate of $q_1$
($q_2$) and $\bar {q}_2$ ($\bar {q}_1$), respectively.
With the wave functions we obtain
\begin{eqnarray}
<q_1\bar {q}_2,q_2\bar {q}_1 \mid H_{\rm I} \mid q_1\bar {q}_1,q_2\bar {q}_2> 
& = & 
\int d^3r_{q_1\bar {q}_1} d^3R_{q_1\bar {q}_1}
     d^3r_{q_2\bar {q}_2} d^3R_{q_2\bar {q}_2}          
    \psi^+_{q_1\bar {q}_2,q_2\bar {q}_1} H_I
    \psi_{q_1\bar {q}_1,q_2\bar {q}_2}       \nonumber    \\
& = & (2\pi)^3\delta (\vec {P}_{\rm f} - \vec {P}_{\rm i})
\frac {{\cal M}_{fi}}{V^2\sqrt {2E_{q_1\bar {q}_1}2E_{q_2\bar {q}_2}
2E_{q_1\bar {q}_2}2E_{q_2\bar {q}_1}}} 
\end{eqnarray}
where $\vec {P}_{\rm i}$ ($\vec {P}_{\rm f}$) is the total three-dimensional 
momentum of the two initial (final) mesons,
and $E_{q_1\bar {q}_1}$, $E_{q_2\bar {q}_2}$, $E_{q_1\bar {q}_2}$ and
$E_{q_2\bar {q}_1}$ are the energies of the four mesons $q_1\bar {q}_1$, 
$q_2\bar {q}_2$, $q_1\bar {q}_2$ and $q_2\bar {q}_1$, respectively. 
The transition amplitude ${\cal M}_{fi}$ is
\begin{eqnarray}
{\cal M}_{fi} & = & \sqrt {2E_{q_1\bar {q}_1}2E_{q_2\bar {q}_2}
2E_{q_1\bar {q}_2}2E_{q_2\bar {q}_1}}
\int d^3r_{q_1\bar {q}_1} d^3r_{q_2\bar {q}_2}
     d^3r_{q_1\bar {q}_1,q_2\bar {q}_2}          \nonumber   \\
& & \psi^+_{q_1\bar {q}_2} (\vec {r}_{q_1\bar {q}_2}) 
\psi^+_{q_2\bar {q}_1} (\vec {r}_{q_2\bar {q}_1}) H_I
\psi_{q_1\bar {q}_1} (\vec {r}_{q_1\bar {q}_1}) 
\psi_{q_2\bar {q}_2} (\vec {r}_{q_2\bar {q}_2})
{\rm e}^{i\vec {p}_{q_1\bar {q}_1,q_2\bar {q}_2} \cdot
\vec {r}_{q_1\bar {q}_1,q_2\bar {q}_2}
-i\vec {p}_{q_1\bar {q}_2,q_2\bar {q}_1} \cdot
\vec {r}_{q_1\bar {q}_2,q_2\bar {q}_1}}    \nonumber \\
\end{eqnarray}
where $\vec {p}_{q_1\bar {q}_1,q_2\bar {q}_2}$ 
($\vec {p}_{q_1\bar {q}_2,q_2\bar {q}_1}$) and
$\vec {r}_{q_1\bar {q}_1,q_2\bar {q}_2}$
($\vec {r}_{q_1\bar {q}_2,q_2\bar {q}_1}$) are the relative momentum and the
relative coordinate of $q_1\bar {q}_1$ ($q_1\bar {q}_2$) and 
$q_2\bar {q}_2$ ($q_2\bar {q}_1$), respectively. $\psi_{ab} (\vec {r}_{ab})$
is the wave function of the relative motion of constituents $a$ and $b$ in 
coordinate space.
Further derivation similar to Ref. \cite{bd} leads to cross section for 
$A(q_{1}\overline{q}_{1})+B(q_{2}\overline{q}_{2})\to
C(q_{1}\overline{q}_{2})+D(q_{2}\overline{q}_{1})$ 
\begin{eqnarray}
\sigma & = & \frac {(2\pi)^4}
{4\sqrt {(P_{q_1\bar {q}_1} \cdot P_{q_2\bar {q}_2})^2 
-m_{q_1\bar {q}_1}^2m_{q_2\bar {q}_2}^2}}    \nonumber   \\
& & \int \frac {d^3P_{q_1\bar {q}_2}}{(2\pi)^32E_{q_1\bar {q}_2}}
     \frac {d^3P_{q_2\bar {q}_1}}{(2\pi)^32E_{q_2\bar {q}_1}}
\mid {\cal M}_{fi} \mid^2     \delta (E_{\rm f} - E_{\rm i})
\delta (\vec {P}_{\rm f} - \vec {P}_{\rm i})      
\end{eqnarray}
where $m_{q_1\bar {q}_1}$ ($m_{q_2\bar {q}_2}$) and
$P_{q_1\bar {q}_1} = (E_{q_1\bar {q}_1}, \vec {P}_{q_1\bar {q}_1})$
($P_{q_2\bar {q}_2} = (E_{q_2\bar {q}_2}, \vec {P}_{q_2\bar {q}_2})$) are the
mass and the four-momentum of meson $A(q_1\bar {q}_1)$ ($B(q_2\bar {q}_2)$),
respectively. Expressed in terms of
the Mandelstam variables $s=(E_{q_1\bar {q}_1}+E_{q_2\bar {q}_2})^2
-(\vec {P}_{q_1\bar {q}_1}+\vec {P}_{q_2\bar {q}_2})^2$ and
$t=(E_{q_1\bar {q}_1}-E_{q_1\bar {q}_2})^2
-(\vec {P}_{q_1\bar {q}_1}-\vec {P}_{q_1\bar {q}_2})^2$, the cross section 
becomes
\begin{equation}
\sigma =\frac{1}{32\pi s}\frac{|\vec{P}^{\prime }(\sqrt{s})|
}{|\vec{P}(\sqrt{s})|}\int_{0}^{\pi }d\theta
|\mathcal{M}_{fi} (s,t)|^{2}\sin \theta ,
\end{equation}
where $\theta $ is the angle between $\vec P$ and $\vec {P}^\prime$,
$\vec {P} = \vec {P}_{q_1\bar {q}_1}= -\vec {P}_{q_2\bar {q}_2}$ 
and $\vec {P}^\prime = \vec {P}_{q_1\bar {q}_2}= -\vec {P}_{q_2\bar {q}_1}$
in the center-of-momentum frame,
\begin{equation}
|\vec {P}(\sqrt{s})|^{2}=\frac{1}{4s}\left\{ \left[ s-\left(
m_{q_1\bar {q}_1}^{2}+m_{q_2\bar {q}_2}^{2}\right) \right]^{2} 
-4m_{q_1\bar {q}_1}^{2}m_{q_2\bar {q}_2}^{2} \right\},
\end{equation}
\begin{equation}
|\vec {P}^{\prime}(\sqrt{s})|^{2}=\frac{1}{4s}\left\{ \left[
s-\left( m_{q_1\bar {q}_2}^2+m_{q_2\bar {q}_1}^2\right)\right]^2
-4m_{q_1\bar {q}_2}^2m_{q_2\bar {q}_1}^2
\right\} ,
\end{equation}%
where $m_{q_1\bar {q}_2}$ and $m_{q_2\bar {q}_1}$ are the masses of the two 
mesons $C(q_{1}\overline{q}_{2})$ and $D(q_{2} \overline{q}_{1})$, 
respectively.
Inelastic reactions are divided into two types: endothermic reactions 
which have $\vec {P} \ne 0$, $\vec {P}' =0$ and $\sigma =0$ at threshold energy
and exothermic reactions which possess $\vec {P} = 0$, $\vec {P}' \ne 0$ 
and $\sigma =+\infty$ at threshold energy.

Since quark or antiquark interchange and gluon exchange are two basic 
processes, ``prior'' diagrams in Fig. 1 and ``post'' diagrams in Fig. 2
are involved in the Born-order meson-meson scattering $A+B \to C+D$.
A scattering in the
prior form means that the gluon exchange takes place prior to quark
or antiquark interchange while the interchange in the post form is
followed by the gluon exchange. The two forms may give different cross sections
for a meson-meson scattering, which is
the so-called post-prior discrepancy \cite{BBS,WC,MM}. 
If the constituent-constituent interaction and the wave function
$\psi_{ab} (\vec {r}_{ab})$ determined by the interaction are used to
calculate the transition amplitude, the post-prior discrepancy disappears, i.e.
the results individually
produced by the post form and the prior form are identical. The
post-prior discrepancy may arise if the interaction used in the transition
amplitude differs from the one used in the determination of 
$\psi_{ab} (\vec {r}_{ab})$ or $\psi_{ab} (\vec {r}_{ab})$ used in the 
transition amplitude is not that mesonic quark-antiquark relative-motion wave
function determined by the Schr$\rm \ddot o$dinger equation.
Cross section for a
meson-meson scattering is assumed to be the average of the two results obtained
in the prior form and in the post form, respectively.
$H_{\rm I}$ in the prior form is
\begin{eqnarray}
H_{\rm I}^{\rm prior} & = & \int \frac {d^3Q}{(2\pi)^3}
V_{q_1\bar {q}_2} (\vec {Q})
{\rm e}^{i\vec {Q} \cdot \vec {r}_{q_1\bar {q}_2}}
+\int \frac {d^3Q}{(2\pi)^3}
V_{\bar {q}_1 q_2} (\vec {Q})
{\rm e}^{i\vec {Q} \cdot \vec {r}_{\bar {q}_1 q_2}}    \nonumber   \\
& & +\int \frac {d^3Q}{(2\pi)^3}
V_{q_1 q_2} (\vec {Q})
{\rm e}^{i\vec {Q} \cdot \vec {r}_{q_1 q_2}}
+\int \frac {d^3Q}{(2\pi)^3}
V_{\bar {q}_1 \bar {q}_2}(\vec {Q})
{\rm e}^{i\vec {Q} \cdot \vec {r}_{\bar {q}_1 \bar {q}_2}}
\end{eqnarray}
where $\vec Q$ is the gluon momentum and
$V_{ab}(\vec {Q})$ is the potential $V_{\rm si}+V_{\rm ss}$ in momentum space,
\begin{eqnarray}
V_{ab}\left( \vec {Q}\right)&=&\frac{ \vec {\lambda }_{a}}{2}
\cdot \frac{\vec {\lambda }_{b}}{2}\frac{16\pi ^{2}}{\vec {Q}^{2}
} \rho \left( \vec {Q}^{2}\right) -\frac{\vec {\lambda }_{a}}{2}
\cdot \frac{\vec {\lambda }_{b}}{2}\frac{16\pi ^{2}}{25}\frac{
\vec {s}_{a}\cdot \vec {s}_{b}}{m_{a}m_{b}}  \notag \\
&&+\frac{\vec {\lambda }_{a}}{2}\cdot \frac{\vec {\lambda }_{b}}{
2}\frac{16\pi ^{2}\lambda }{25Q}\int_{0}^{+\infty}dx\frac{d^{2}v\left(
x\right) }{dx^{2}}\sin \left( \frac{Q}{\lambda }x\right) 
\frac{\vec {s}_{a}\cdot \vec {s}_{b}}{m_{a}m_{b}},
\end{eqnarray}
The transition amplitude in the prior form is
\begin{eqnarray}
{\cal M}_{fi}^{\rm prior} & = &
\sqrt {2E_{q_1\bar {q}_1}2E_{q_2\bar {q}_2}2E_{q_1\bar {q}_2}
2E_{q_2\bar {q}_1}}
\int \frac {d^3 p_{q_1\bar {q}_2}}{(2\pi)^3} 
     \frac {d^3 p_{q_2\bar {q}_1}}{(2\pi)^3}      \nonumber    \\
& & \psi^+_{q_1\bar {q}_2} (\vec {p}_{q_1\bar {q}_2}) 
\psi^+_{q_2\bar {q}_1} (\vec {p}_{q_2\bar {q}_1})
(V_{q_1\bar {q}_2}+V_{\bar {q}_1 q_2}+V_{q_1 q_2}+V_{\bar {q}_1 \bar {q}_2})
\psi_{q_1\bar {q}_1} (\vec {p}_{q_1\bar {q}_1}) 
\psi_{q_2\bar {q}_2} (\vec {p}_{q_2\bar {q}_2})    \nonumber   \\
\end{eqnarray}
where $\psi_{ab} (\vec {p}_{ab})$ is the wave function of the relative motion
of constituents $a$ and $b$ in momentum space and satisfies
$\int \frac {d^3p_{ab}}{(2\pi)^3} \psi^+_{ab} (\vec {p}_{ab})
\psi_{ab} (\vec {p}_{ab}) =1$. $\vec {p}_{ab}$ is the relative momentum of $a$
and $b$. $H_{\rm I}$ in the post form is
\begin{eqnarray}
H_{\rm I}^{\rm post} & = & \int \frac {d^3Q}{(2\pi)^3}
V_{q_1\bar {q}_1}(\vec {Q})
{\rm e}^{i\vec {Q} \cdot \vec {r}_{q_1\bar {q}_1}}
+\int \frac {d^3Q}{(2\pi)^3}
V_{\bar {q}_2 q_2}(\vec {Q})
{\rm e}^{i\vec {Q} \cdot \vec {r}_{\bar {q}_2 q_2}}   \nonumber  \\ 
& & +\int \frac {d^3Q}{(2\pi)^3}
V_{q_1 q_2}(\vec {Q})
{\rm e}^{i\vec {Q} \cdot \vec {r}_{q_1 q_2}}
+\int \frac {d^3Q}{(2\pi)^3}
V_{\bar {q}_1 \bar {q}_2}(\vec {Q})
{\rm e}^{i\vec {Q} \cdot \vec {r}_{\bar {q}_1 \bar {q}_2}}
\end{eqnarray}
which leads to the transition amplitude in the post form
\begin{eqnarray}
{\cal M}_{fi}^{\rm post} & = & 
\sqrt {2E_{q_1\bar {q}_1}2E_{q_2\bar {q}_2}2E_{q_1\bar {q}_2}
2E_{q_2\bar {q}_1}}        \nonumber   \\
& & ( \int \frac {d^3 p_{q_1\bar {q}_1}}{(2\pi)^3} 
     \frac {d^3 p_{q_1\bar {q}_2}}{(2\pi)^3}    
\psi^+_{q_1\bar {q}_2} (\vec {p}_{q_1\bar {q}_2}) 
\psi^+_{q_2\bar {q}_1} (\vec {p}_{q_2\bar {q}_1})
V_{q_1\bar {q}_1}
\psi_{q_1\bar {q}_1} (\vec {p}_{q_1\bar {q}_1}) 
\psi_{q_2\bar {q}_2} (\vec {p}_{q_2\bar {q}_2})       \nonumber   \\
& & + \int \frac {d^3 p_{q_2\bar {q}_2}}{(2\pi)^3} 
           \frac {d^3 p_{q_2\bar {q}_1}}{(2\pi)^3}    
\psi^+_{q_1\bar {q}_2} (\vec {p}_{q_1\bar {q}_2}) 
\psi^+_{q_2\bar {q}_1} (\vec {p}_{q_2\bar {q}_1})
V_{\bar {q}_2q_2}
\psi_{q_1\bar {q}_1} (\vec {p}_{q_1\bar {q}_1}) 
\psi_{q_2\bar {q}_2} (\vec {p}_{q_2\bar {q}_2})       \nonumber   \\
& & + \int \frac {d^3 p_{q_1\bar {q}_2}}{(2\pi)^3} 
           \frac {d^3 p_{q_2\bar {q}_1}}{(2\pi)^3}    
\psi^+_{q_1\bar {q}_2} (\vec {p}_{q_1\bar {q}_2}) 
\psi^+_{q_2\bar {q}_1} (\vec {p}_{q_2\bar {q}_1})
(V_{q_1 q_2}+V_{\bar {q}_1 \bar {q}_2})
\psi_{q_1\bar {q}_1} (\vec {p}_{q_1\bar {q}_1}) 
\psi_{q_2\bar {q}_2} (\vec {p}_{q_2\bar {q}_2})  )     \nonumber   \\
\end{eqnarray}

Let $\sigma (S,m_S,\sqrt {s})\equiv \sigma$ 
denote the cross section corresponding to a 
component $m_S$ of the total spin $S$ of the two incoming mesons 
in the following eleven channels of 
$A(q_{1}\overline{q}_{1})+B(q_{2}\overline{q}_{2}) \to
C(q_{1}\overline{q}_{2})+D(q_{2}\overline{q}_{1})$
\begin{displaymath}
(1)~q_{1}\overline{q}_{1}(S_{A}=0)+q_{2}\overline{q}_{2}(S_{B}=0)\rightarrow
q_{1}\overline{q}_{2}(S_{C}=0)+q_{2}\overline{q}_{1}(S_{D}=0);~S=0;
\end{displaymath}
\begin{displaymath}
(2)~q_{1}\overline{q}_{1}(S_{A}=0)+q_{2}\overline{q}_{2}(S_{B}=0)\rightarrow
q_{1}\overline{q}_{2}(S_{C}=1)+q_{2}\overline{q}_{1}(S_{D}=1);~S=0;
\end{displaymath}
\begin{displaymath}
(3)~q_{1}\overline{q}_{1}(S_{A}=0)+q_{2}\overline{q}_{2}(S_{B}=1)\rightarrow
q_{1}\overline{q}_{2}(S_{C}=0)+q_{2}\overline{q}_{1}(S_{D}=1);~S=1;
\end{displaymath}
\begin{displaymath}
(4)~q_{1}\overline{q}_{1}(S_{A}=0)+q_{2}\overline{q}_{2}(S_{B}=1)\rightarrow
q_{1}\overline{q}_{2}(S_{C}=1)+q_{2}\overline{q}_{1}(S_{D}=0);~S=1;
\end{displaymath}
\begin{displaymath}
(5)~q_{1}\overline{q}_{1}(S_{A}=0)+q_{2}\overline{q}_{2}(S_{B}=1)\rightarrow
q_{1}\overline{q}_{2}(S_{C}=1)+q_{2}\overline{q}_{1}(S_{D}=1);~S=1;
\end{displaymath}
\begin{displaymath}
(6)~q_{1}\overline{q}_{1}(S_{A}=1)+q_{2}\overline{q}_{2}(S_{B}=1)\rightarrow
q_{1}\overline{q}_{2}(S_{C}=0)+q_{2}\overline{q}_{1}(S_{D}=0);~S=0;
\end{displaymath}
\begin{displaymath}
(7)~q_{1}\overline{q}_{1}(S_{A}=1)+q_{2}\overline{q}_{2}(S_{B}=1)\rightarrow
q_{1}\overline{q}_{2}(S_{C}=0)+q_{2}\overline{q}_{1}(S_{D}=1);~S=1;
\end{displaymath}
\begin{displaymath}
(8)~q_{1}\overline{q}_{1}(S_{A}=1)+q_{2}\overline{q}_{2}(S_{B}=1)\rightarrow
q_{1}\overline{q}_{2}(S_{C}=1)+q_{2}\overline{q}_{1}(S_{D}=0);~S=1;
\end{displaymath}
\begin{displaymath}
(9)~q_{1}\overline{q}_{1}(S_{A}=1)+q_{2}\overline{q}_{2}(S_{B}=1)\rightarrow
q_{1}\overline{q}_{2}(S_{C}=1)+q_{2}\overline{q}_{1}(S_{D}=1);~S=0;
\end{displaymath}
\begin{displaymath}
(10)~q_{1}\overline{q}_{1}(S_{A}=1)+q_{2}\overline{q}_{2}(S_{B}=1)\rightarrow
q_{1}\overline{q}_{2}(S_{C}=1)+q_{2}\overline{q}_{1}(S_{D}=1);~S=1;
\end{displaymath}
\begin{displaymath}
(11)~q_{1}\overline{q}_{1}(S_{A}=1)+q_{2}\overline{q}_{2}(S_{B}=1)\rightarrow
q_{1}\overline{q}_{2}(S_{C}=1)+q_{2}\overline{q}_{1}(S_{D}=1);~S=2;
\end{displaymath}
where the quantities $S_A$, $S_B$, $S_C$ and $S_D$ in the parentheses are meson
spins. In fact, $\sigma (S,m_S,\sqrt {s})$ is independent of $m_S$ and is
calculated at any value subject to the condition $-S \leq m_S \leq S$.
When the orbital angular momenta of the four mesons 
$A$, $B$, $C$ and $D$ are all zero, the unpolarized cross section is
\begin{equation}
\sigma ^{\rm unpol} (\sqrt {s})=
\frac{1}{(2S_{A}+1)(2S_{B}+1)}\sum _{S}(2S+1)\sigma(S,m_S,\sqrt {s})
\end{equation}
where $S$ takes values that are allowed by $\mid S_A -S_B \mid \leq S \leq
S_A +S_B$ and $\mid S_C -S_D \mid \leq S \leq S_C +S_D$.

\vspace{0.5cm}
\leftline {\bf 4. Numerical results and discussions}
\vspace{0.5cm}

The transition amplitude $\mathcal{M}_{fi}$ is obtained from
meson wave functions of which each is the product
of quark-antiquark relative-motion, spin, flavor and color wave functions.
The quark-antiquark relative-motion wave function $\psi_{ab}(\vec {p}_{ab})$
is the Fourier transform of the wave function $\psi_{ab}(\vec {r}_{ab})$ what
is obtained by the Schr$\rm \ddot o$dinger equation with the 
Buchm$\rm \ddot u$ller-Tye potential. 
$\psi_{q_1\bar {q}_1}(\vec {p}_{q_1\bar {q}_1})$,
$\psi_{q_2\bar {q}_2}(\vec {p}_{q_2\bar {q}_2})$,
$\psi_{q_1\bar {q}_2}(\vec {p}_{q_1\bar {q}_2})$ and
$\psi_{q_2\bar {q}_1}(\vec {p}_{q_2\bar {q}_1})$ take the same form.
The matrix elements of $\vec {s}_a \cdot \vec {s}_b$ are derived for the
eleven channels. The transition amplitude is calculated in the 
center-of-momentum frame of the two initial mesons where
\begin{equation}
4\sqrt{E_{q_1\bar {q}_1}E_{q_2\bar {q}_2}E_{q_1\bar {q}_2}E_{q_2\bar {q}_1}}
=\frac{1}{s}\sqrt{
[s^{2}-(m_{q_1\bar {q}_1}^2-m_{q_2\bar {q}_2}^{2})^2]
[s^{2}-(m_{q_1\bar {q}_2}^{2}-m_{q_2\bar {q}_1}^{2})^{2}]}.
\end{equation}

\vspace{0.25cm}
\leftline{4.1. Elastic phase shifts for $I=2~\pi \pi $ and $I=3/2~K \pi $ 
scattering} 
\vspace{0.25cm}

The phase shift formula is \cite{BSW92}
\begin{equation}\label{ps}
\delta_{l}=-\frac{2\pi ^{2}|\vec {P}|
E_{q_1\bar {q}_1}E_{q_2\bar {q}_2}}{E_{q_1\bar {q}_1}+E_{q_2\bar {q}_2}}
\int_{-1}^{1}T_{fi}P_{l}(x^\prime )dx^\prime,
\end{equation}
where $P_l (x^\prime )$ with $x^\prime =\cos \theta $ is the Legendre 
polynomial. $T_{fi}$ is related to
the transition amplitudes in the prior form and in the post form
\begin{equation}
T_{fi}=\frac {1}{(2\pi)^3
\sqrt{2E_{q_1\bar {q}_1}2E_{q_2\bar {q}_2}2E_{q_1\bar {q}_2}
2E_{q_2\bar {q}_1}}}
\frac {\mathcal{M}_{fi}^{\rm prior}+\mathcal{M}_{fi}^{\rm post}}{2}
\end{equation}

The S-wave elastic phase shifts for $I=2~\pi \pi $ and
$I=3/2~K \pi $ scattering given by Eq.~(\ref{ps}) are
plotted as solid curves in Figs.~\ref{fig3} and \ref{fig4}, respectively. 
Our theoretical values for the $I=2~\pi \pi $ elastic scattering are in
good agreement with the experimental data. Our theoretical values for
the $I=3/2~K \pi$ cannot completely match the poor experimental data but agree
with from threshold energy to $\sqrt {s} = 1.3~\rm{GeV}$ what is the region 
accessible in hadronic matter. 

The solid curve is obtained when physical meson masses and flavor-symmetry
breaking are used and the quark-antiquark relative-motion wave functions of
$K$ and $\pi$ are taken to be identical. The nonstrange and strange quark 
masses remain as the values determined by ground-state
meson spectroscopy via 
the Schr$\rm \ddot o$dinger equation with the Buchm$\rm \ddot u$ller-Tye
potential in Section 2. Squares and circles in Fig. 4 are data of Ref. 
\cite{Jon73} and Ref. \cite{Est78}, respectively. The two measurements may
give different data when the center-of-mass energy of $K$ and $\pi$ is 
larger than 1.2 GeV. The possible discrepancy requires further examination of
$K\pi$ phase shift in both experiment and theory. This is one reason why we
calculate the $I=3/2$ $K\pi$ phase shift.

The transition amplitude is calculated with the quark-antiquark relative-motion
wave functions what are solutions of the Schr$\rm \ddot o$dinger equation with 
the central spin-independent potential. But the interaction employed in the
transition amplitude contains both the central spin-independent potential
and the spin-spin term. Then the wave functions are not exact with respect to
the interaction. Therefore, the post-prior discrepancy should come from the
approximate wave functions and the spin-spin term. Fortunately, in the elastic
scatterings where meson masses do not change, the spin-spin term gives almost 
the same magnitudes to the transition amplitudes individually obtained
in the post form and in the
prior form, and the post-prior discrepancy is completely negligible. For
inelastic scatterings where meson masses change, the spin-spin term produces
different magnitudes, hence the post-prior discrepancy exists.

\vspace{0.25cm}
\leftline{4.2. Cross sections for inelastic scatterings of nonstrange mesons}
\vspace{0.25cm}

Unpolarized cross sections for the $I=2~\pi \pi \to \rho \rho $ and $I=2~\rho
\rho \to \pi \pi $ are shown in Fig.~\ref{fig5}. The former has a
maximum cross section of $0.47~\rm{mb}$ at $\sqrt{s}=1.79~\rm{GeV}$. 
Since no $s$-channel resonances contribute significantly to the reaction
$\pi^+ \pi^- \leftrightarrow \omega \omega$ as well as no quark-antiquark 
annihilation happens, 
we can use the quark-interchange mechanism to calculate the cross 
section for the reaction $\pi^+ \pi^- \leftrightarrow \omega \omega$.
Flavor matrix elements for the $I=2~\pi
\pi \leftrightarrow \rho \rho $ and $\pi^+ \pi^- \leftrightarrow
\omega \omega $ inelastic scatterings are $1$ and $-1/2$,
respectively. The discrepancy of the cross sections for the two reactions 
comes from their
different flavor matrix elements while their color and spin matrix elements
are not different and the integration over relative-momentum variables in
${\cal M}_{fi}^{\rm prior}$ and ${\cal M}_{fi}^{\rm post}$ with nearly equal 
$\rho$ and $\omega$ masses offers almost the same values. 
Then the unpolarized cross section for the reaction 
$\pi^+ \pi^- \leftrightarrow \omega \omega$ is one fourth of the one 
for the reaction of $I=2~\pi \pi \leftrightarrow \rho \rho$. 

Unpolarized cross sections for the $I=2~\pi \rho \to \rho \rho $ and $I=2~\rho
\rho \to \pi \rho $ are shown in Fig.~\ref{fig6}. The endothermic reaction has 
a maximum cross section of $0.73~\rm{mb}$ at $\sqrt{s}=1.86~\rm{GeV}$. 
The quark-interchange mechanism can also be applied to calculate cross sections
for $\pi^+ \rho^- \leftrightarrow \omega \omega$ and 
$\pi^- \rho^+ \leftrightarrow \omega \omega$ inelastic scatterings.
According to flavor matrix elements, the unpolarized cross section for the
reaction $\pi^+ \rho^- \leftrightarrow \omega \omega$ or
$\pi^- \rho^+ \leftrightarrow \omega \omega$ is one fourth of the one for the 
reaction of $I=2~\pi \rho \leftrightarrow \rho \rho $. The two reactions of
$I=2~\pi \pi \to \rho \rho $ and
$I=2~\pi \rho \to \rho \rho $ have the same
threshold energy but the latter has larger cross section than the former.
Such difference originates from the physical masses of initial mesons 
and the total spins $S=0$ for the $I=2~\pi \pi \to \rho \rho $ and
$S=1$ for the $I=2~\pi \rho \to \rho \rho $.

\vspace{0.25cm}
\leftline{4.3. Cross sections for inelastic scatterings of strange mesons}
\vspace{0.25cm}

Unpolarized cross sections for the $I=1~K K \leftrightarrow K^\ast K^\ast $ and
$I=1~K K^\ast \leftrightarrow K^\ast K^\ast $ inelastic scatterings
are shown in Figs.~\ref{fig7} and \ref{fig8}, respectively. Since the two
reactions have the same threshold energy, they can be compared.
The reaction of $I=1~K K \to K^\ast K^\ast $ has a maximum cross section of 
$0.60~\rm{mb}$ at $\sqrt{s}=1.94~\rm{GeV}$ while the reaction of 
$I=1~K K^\ast \to
K^\ast K^\ast $  has a maximum cross section of $0.84~\rm{mb}$ at
$\sqrt{s}=2.09~\rm{GeV}$. Given a center-of-mass energy $\sqrt s$, according 
to Eqs. (15) and (16) the outgoing $K^*$ mesons of the two reactions
have the same momentum $\mid \vec {P}^\prime \mid$, but the $K$ mesons in the
reaction of $I=1~K K \to K^\ast K^\ast $ have larger momentum 
$\mid \vec {P} \mid$ than the initial $K^*$ and $K$ mesons in the reaction of
$I=1~K K^\ast \to K^\ast K^\ast $. Then the factor
$\mid \vec {P}^\prime \mid / \mid \vec {P} \mid$ what has a smaller value for 
the former than the one for the latter gives a smaller cross section to the
former than the latter in spite of the difference generated by the different 
total spins of the two reactions. Among inelastic scatterings of strange 
mesons,
the reactions $K K \leftrightarrow K K^\ast$ are forbidden since the total spin
of the two incoming mesons is not equal to that of the two outgoing mesons.

\vspace{0.25cm}
\leftline{4.4. Cross sections for inelastic scatterings of nonstrange mesons 
by strange mesons}
\vspace{0.25cm}

This subsection contributes to the $I=3/2$ inelastic scatterings between one
of $\pi$ and $\rho$ and one of $K$ and $K^*$. 
Figs.~\ref{fig9}, \ref{fig10} and \ref{fig11} individually show unpolarized
cross sections for the inelastic scatterings of
$I=3/2~\pi K \leftrightarrow \rho K^\ast$, $I=3/2~\pi K^\ast
\leftrightarrow \rho K^\ast$ and $I=3/2~\rho K \leftrightarrow \rho K^\ast $ 
which have the same threshold energy.
The reaction of $I=3/2~\pi K \to \rho K^\ast$ ($I=3/2~\pi K^\ast \to \rho
 K^\ast$, $I=3/2~\rho K \to \rho K^\ast $) has a maximum cross
section of $0.52~\rm{mb}$ ($0.47~\rm{mb}$, $0.51~\rm{mb}$) at
$\sqrt{s}=1.76~\rm{GeV}$ ($1.94~\rm{GeV}$, $1.99~\rm{GeV}$). 
The maximum cross sections for the three reactions are comparable to those 
shown in the last two subsections. Attractively,
the reaction of $I=3/2~\pi K^\ast \to \rho K$ exhibited in Fig. 12 
has a maximum cross section of $1.41~\rm{mb}$ at $\sqrt{s}=1.42~\rm{GeV}$.
The maximum cross section for this reaction is quite lager that any maximum
cross section seen before. This case is caused by the small threshold energy 
1.266 GeV and the samll difference 0.234 GeV of the mass sums of the final
mesons and of the initial mesons. Therefore, the reaction of
$I=3/2~\pi K^\ast \to \rho K$
is most important among the endothermic nonresonant reactions.

\vspace{0.25cm}
\leftline{4.5. Cross sections for $\phi$ meson in collisions with $\pi$ and 
$\rho$ mesons} 
\vspace{0.25cm}

Unpolarized cross sections for the $I=1~\pi \phi \to K K^\ast$
(or $K^\ast K$) and the $I=1~\pi \phi \to K^\ast K^\ast $ are shown in
Figs.~\ref{fig13} and \ref{fig14}, respectively. $K$ denotes either
a kaon or an antikaon as appropriate. These reactions are endothermic. 
The reaction of $I=1~\pi \phi \to K K^\ast$ (or $K^\ast K$) has a
maximum cross section of $0.99~\rm{mb}$ at $\sqrt{s}=1.51~\rm{GeV}$ near the
threshold energy. The cross section for the reaction of 
$I=1~\pi \phi \to K^\ast K^\ast $ features a wide shape with a maximum 
of $0.65~\rm{mb}$ at $\sqrt{s}=2.11~\rm{GeV}$. The exothermic reaction
$\pi \phi \to K K$ is forbidden since the total spin of the two incoming
mesons does not equal that of the two outgoing mesons.

Unpolarized cross sections for the reactions of $I=1~\rho \phi \to K K$, 
$I=1~\rho \phi \to K K^\ast$ (or $K^\ast K$) and 
$I=1~\rho \phi \to K^\ast K^\ast$ 
are shown in the upper, middle and lower panels in Fig.~\ref{fig15}, 
respectively. These exothermic reactions have the same threshold energy which 
equals the sum of the physical $\rho$ and $\phi$ masses. The cross sections
for the $I=1~\rho \phi \to K K$ and the $I=1~\rho \phi \to K K^\ast$ (or
$K^\ast K$) decreases very rapidly from the threshold energy but the cross
section for the $I=1~\rho \phi \to K^\ast K^\ast$ decreases slowly from 
$\sqrt {s}=2$ GeV.

Some calculations of transport equations utilize a constant cross section for 
the scattering of $\phi$ and a meson as an input. A choice on the value of the 
constant cross section lacks of direct experimental and theoretical support. 
To guide the choice, we evaluate the
constant from the meson-$\phi$ cross section for a reaction that we have
studied so far. We define the average cross section
for $\phi$ in collision with a meson
\begin{equation}
\langle \sigma_{m\phi}\rangle
=\frac{g_m \int\frac{d^3k_1}{(2\pi)^3}f_{m}(k_1)
\frac{d^3k_2}{(2\pi)^3}f_{\phi}(k_2)\sigma_{m\phi}(\sqrt{s})}
{g_m\int\frac{d^3k_1}{(2\pi)^3}f_{m}(k_1)
\int\frac{d^3k_2}{(2\pi)^3}f_{\phi}(k_2)},
\end{equation}
where $g_m$ is the spin degeneracy factor of the meson,  
$f_{m}$ and $f_{\phi}$ are the momentum distributions of the meson and
$\phi$, respectively; $\sigma_{m\phi}(\sqrt{s})$ is the $\sqrt s$-dependent
meson-$\phi$ cross section.
$\langle \sigma_{m\phi}\rangle$ is independent of
$\sqrt s$ that relates to $k_1$ and $k_2$ and is suggested as the constant.

\begin{table}[htbp]
\centering \caption{Average values for $\phi$ meson in
collisions with $\pi$ and $\rho$ mesons.}
\label{avercross}
\begin{tabular*}{12cm}{@{\extracolsep{\fill}}ccc}
  \hline
  Channel & $\langle\sigma_{m\phi}\rangle~(\rm{mb})$  
 & $\langle v_{\rm rel}\sigma_{m\phi}\rangle~(\rm{mb})$\\
  \hline
  $I=1~\pi \phi \to K K^\ast$ (or $K^\ast K$) & 0.365 & 0.374 \\
  $I=1~\pi \phi \to K^\ast K^\ast $ & 0.056 & 0.073 \\
  $I=1~\rho \phi \to K K$ & 0.590 & 0.248 \\
  $I=1~\rho \phi \to K K^\ast$ (or $K^\ast K$) & 0.624 & 0.284 \\
  $I=1~\rho \phi \to K^\ast K^\ast$ & 2.162 & 1.235 \\
  \hline
\end{tabular*}
\end{table}

Since $\phi$ mesons in hadronic matter may be in thermal equilibrium 
\cite{AK,greene},
we assume $f_m(k_1) = e^{-E_m/T}$ with the meson energy $E_m$ and 
$f_{\phi}(k_2) = e^{-E_{\phi}/T}$ with the $\phi$ energy $E_\phi$. 
At temperature $T=0.15~\mathrm{GeV}$, average cross sections 
are listed in Table \ref{avercross}. The results for 
the reactions of $I=1~\pi \phi \to K^\ast K^\ast$ and $I=1~\rho \phi 
\to K^\ast K^\ast $ need to be specified. Since the difference of the 
total masses of the initial mesons and of the final mesons in the
$I=1~\pi \phi \to K^\ast K^\ast $ reaction is 0.63 GeV, what is much larger 
than the
temperature, the momenta of $\pi$ and $\phi$ that can trigger the reaction
are at least 0.59 GeV much larger than $T$. 
Then $f_m \ll 1$ and $f_{\phi} \ll 1$ leads to the 
very small average cross section 0.056 mb.
Since the unpolarized cross section for the $I=1~\rho \phi \to K^\ast K^\ast$
decreases slowly in the region that is not far away from the threshold energy,
$f_m$ and $f_\phi$ at low momenta of $\rho$ and $\phi$ produce the 
particularly large average cross section for the $I=1~\rho \phi 
\to K^\ast K^\ast$, which indicates that the reaction is important. 

In the last column is listed another interesting thermal-averaged quantity
\begin{equation}
\langle v_{\rm rel} \sigma_{m\phi}\rangle
=\frac{g_m \int\frac{d^3k_1}{(2\pi)^3}f_{m}(k_1)
\frac{d^3k_2}{(2\pi)^3}f_{\phi}(k_2)v_{\rm rel}\sigma_{m\phi}(\sqrt{s})}
{g_m\int\frac{d^3k_1}{(2\pi)^3}f_{m}(k_1)
\int\frac{d^3k_2}{(2\pi)^3}f_{\phi}(k_2)},
\end{equation}
where $v_{\rm rel}$ is the relative velocity of the meson and $\phi$. The
orders of $\langle v_{\rm rel}\sigma_{m\phi}\rangle$ are identical with the 
orders of $\langle \sigma_{m\phi}\rangle$. The scales 
$\langle \sigma_{m\phi}\rangle = 0.5~{\rm mb}$ and 
$\langle v_{\rm rel} \sigma_{m\phi}\rangle = 0.3~{\rm mb}$ can be useful in
spite of the types of reactions.

\vspace{0.25cm}
\leftline{4.6. General discussions}
\vspace{0.25cm}

Maximum cross sections for ten endothermic reactions, corresponding 
center-of-mass energy $\sqrt s$, initial meson momentum $\mid \vec {P} \mid$,
final meson momentum $\mid \vec {P}^\prime \mid$ and $\frac {1}{s} 
\frac {\mid \vec {P}^\prime \mid }{\mid \vec {P} \mid }$ are listed
in Table 2. The reactions but the $I=\frac {3}{2}~\pi K^\ast \to \rho K$ and 
the $I=1~\pi \phi \to K K^\ast$ have small differences of the initial meson 
momenta. If the final meson momentum is large (small), the factor
$\frac {1}{s} \frac {\mid \vec {P}^\prime \mid}{\mid \vec {P} \mid }$
is large (small), but the squared transition amplitude $\mid \mathcal 
{M}_{fi} \mid^2$ is small (large) since $\mid \mathcal {M}_{fi} \mid^2$
decreases with increasing $\mid \vec {P}^\prime \mid$. The product of
$\frac {1}{s} \frac {\mid \vec {P}^\prime \mid}{\mid \vec {P} \mid }$
and $\mid \mathcal {M}_{fi} \mid^2$ in the cross section formula (Eq. (14))
changes slowly with respect to the variation of final meson momentum from one 
to another reaction. We can thus understand why the eight endothermic reactions
have similar maximum cross sections, i.e. from 0.45 mb to 0.85 mb. Compared
to the eight reactions, the $I=\frac {3}{2}~\pi K^\ast \to \rho K$ and 
the $I=1~\pi \phi \to K K^\ast$ have larger 
$\frac {1}{s} \frac {\mid \vec {P}^\prime \mid}{\mid \vec {P} \mid }$
and smaller $\mid \vec {P} \mid$ and $\mid \vec {P}^\prime \mid$ which leads to
larger $\mid \mathcal {M}_{fi} \mid^2$. This accounts for the larger cross 
sections of the two reactions than the eight reactions.

\begin{table}[htbp]
\centering \caption{Maximum cross sections $\sigma_{\rm max}$ and 
corresponding variables.}
\label{maxcross}
\begin{tabular*}{16.5cm}{@{\extracolsep{\fill}}cccccc}
  \hline
  reaction & $\sqrt {s}~({\rm GeV})$ 
 & $\sigma_{\rm max}~({\rm mb})$ & $\mid \vec {P} \mid ({\rm GeV})$ 
 & $\mid \vec {P}^\prime \mid ({\rm GeV})$ & $\frac {1}{s} 
  \frac {\mid \vec {P}^\prime \mid }{\mid \vec {P} \mid }~({\rm GeV}^{-2})$  \\
  \hline
  $I=2~\pi \pi \to \rho \rho$ & 1.79 & 0.47 & 0.884 & 0.456 & 0.161\\
  $I=2~\pi \rho \to \rho \rho$ & 1.86 & 0.73 & 0.763 & 0.522 & 0.197\\
  $I=1~KK \to K^\ast K^\ast$ & 1.94 & 0.60 & 0.834 & 0.377 & 0.120 \\
  $I=1~K K^\ast \to K^\ast K^\ast$ & 2.09 & 0.84 & 0.766 & 0.541 & 0.162 \\
  $I=\frac {3}{2}~\pi K \to \rho K^\ast$  
& 1.76 & 0.52 & 0.804 & 0.286 & 0.115 \\
  $I=\frac {3}{2}~\pi K^\ast \to \rho K^\ast$  
& 1.94 & 0.47 & 0.756 & 0.498 & 0.175 \\
  $I=\frac {3}{2}~\rho K \to \rho K^\ast $ 
& 1.99 & 0.51 & 0.760 & 0.545 & 0.181 \\
  $I=\frac {3}{2}~\pi K^\ast \to \rho K$  
& 1.42 & 1.41 & 0.413 & 0.316 & 0.379 \\
  $I=1~\pi \phi \to K K^\ast$ & 1.51 & 0.99 & 0.394 & 0.285 & 0.318 \\
  $I=1~\pi \phi \to K^\ast K^\ast$ & 2.11 & 0.65 & 0.801 & 0.560 & 0.157 \\
  \hline
\end{tabular*}
\end{table}

When $\sqrt s$ increases, $\int^\pi_0 d\theta \sin \theta \mid \mathcal 
{M}_{fi} \mid^2$ decreases slowly near the threshold energy and rapidly in
the other region, and by contrast the factor 
$\frac {1}{s} \frac {\mid \vec {P}^\prime \mid}{\mid \vec {P} \mid }$
increases very rapidly and decreases moderately. The peak of 
$\frac {1}{s} \frac {\mid \vec {P}^\prime \mid}{\mid \vec {P} \mid }$
locates in the slowly-changing region of $\int^\pi_0 d\theta \sin \theta \mid 
\mathcal {M}_{fi} \mid^2$. In case 
$\int^\pi_0 d\theta \sin \theta \mid \mathcal 
{M}_{fi} \mid^2$ for an endothermic reaction decreases very slowly near the
threshold energy, the cross section for the reaction takes its maximum at the
same $\sqrt s$ as 
$\frac {1}{s} \frac {\mid \vec {P}^\prime \mid}{\mid \vec {P} \mid }$
does. Otherwise the peak of cross section locates at the left of but near the
peak of
$\frac {1}{s} \frac {\mid \vec {P}^\prime \mid}{\mid \vec {P} \mid }$. 
Therefore, the energy where the maximum of cross section occurs is
mainly determined by the maximum of  
$\frac {1}{s} \frac {\mid \vec {P}^\prime \mid}{\mid \vec {P} \mid }$.

While the center-of-mass energy $\sqrt s$ exceeds the values shown in Table 2,
cross sections decrease. It is interesting to evaluate the intrinsic momenta
of quark-antiquark relative motion at high energies, for example,
where the cross section is one fourth of the
maximum cross section. The average of the quark-antiquark relative momenta of 
the four mesons in an endothermic reaction is listed in Table 3. If the total 
mass of final mesons is large, the average relative momentum is large. The
differences of the average relative momenta of the reactions in Table 3 are
not large. The average relative momenta stay in the rapid falling region of the
relative-motion wave functions.

\begin{table}[htbp]
\centering \caption{Average quark-antiquark relative momentum.} 
\label{averelm}
\begin{tabular*}{13cm}{@{\extracolsep{\fill}}ccc}
  \hline
  reaction & $\sqrt {s}~({\rm GeV})$  & average relative momentum ($\rm GeV$)\\
  \hline
  $I=2~\pi \pi \to \rho \rho$ & 2.347 & 0.491 \\
  $I=2~\pi \rho \to \rho \rho$ & 2.363 & 0.473 \\
  $I=1~KK \to K^\ast K^\ast$ & 2.703 & 0.580 \\
  $I=1~K K^\ast \to K^\ast K^\ast$ & 2.712 & 0.570 \\
  $I=\frac {3}{2}~\pi K \to \rho K^\ast$  & 2.411 & 0.545 \\
  $I=\frac {3}{2}~\pi K^\ast \to \rho K^\ast$ & 2.513 & 0.550 \\
  $I=\frac {3}{2}~\rho K \to \rho K^\ast $ & 2.507 & 0.553 \\
  $I=\frac {3}{2}~\pi K^\ast \to \rho K$  & 1.796 & 0.445 \\
  \hline
\end{tabular*}
\end{table}

We study meson-meson scattering $A+B \to C+D$ in the Born approximation. The
interaction that affects the scattering is between $A$ constituents and $B$
constituents or between $C$ and $D$. The former must be followed by constituent
rearrangement, i.e. the scattering is described by the ``prior'' diagrams of
Fig. 1. The latter must be preceded by constituent rearrangement, i.e. the 
scattering is described by the ``post'' diagrams of Fig. 1. In principle, we 
may use either the ``prior'' diagrams or the ``post'' diagrams to study the 
process $A+B \to C+D$. But practical calculations may infer
the post-prior discrepancy. 
In this case the prior form is first used to get a result and next
the post form is used to get another result. The average of the two independent
results is taken to describe the process $A+B \to C+D$. The average in fact
avoids double counting in the use of the prior form and the post form, and
reconcile the two independent results. For elastic scatterings where the 
post-prior discrepancy is negligible, we can only employ the post form in
comparison to 
antisymmetrizing the initial state in the resonating group 
method \cite{Zhao98} and generator coordinate method \cite{BDPK}. For inelastic
scattering antisymmetrizing the final state in the resonating group 
method and generator coordinate method is also necessary since the interaction 
exchange kernel resulted from antisymmetrizing the final state probably 
differs from the one resulted from antisymmetrizing the initial state. 
The post and prior forms are related to
antisymmetrizing the initial and final states, respectively.

The relationship of the Born series for elastic scattering 
to the resonating group 
method or the generator coordinate method \cite{BDPK} was detailed by Barnes  
and Swanson \cite{BS92,BS94}. Born-order $T$ matrix is the interaction exchange
kernel in the first term of the iterative procedure of the resonating group
method. The exchange kernel for the pure quark exchange with no gluon exchange 
is proportional to the normalization kernel. It is damped when the 
cluster-cluster distance approaches infinity and hence does not contribute to  
scattering \cite{BS94,OY}. 
The Born approximation does not provide a term corresponding to
the exchange  kernel for the pure quark exchange with no gluon exchange.

Distinct roles of the inelastic scatterings of ground-state mesons can be
identified. For instance, $\pi$ absorption reactions are 
$\pi \pi \to \rho \rho$, $\pi \rho \to \rho \rho$, $\pi K \to \rho K^\ast$, 
$\pi K^\ast \to \rho K^\ast$, $\pi K^\ast \to \rho K$, 
$\pi \phi \to K K^\ast$ (or $K^\ast K$) and $\pi \phi \to K^\ast K^\ast$;   
$K^\ast$ creation reactions are 
$KK \to K^\ast K^\ast$, $K K^\ast \to K^\ast K^\ast$, $\pi K \to \rho K^\ast$, 
$\rho K \to \rho K^\ast$, $\rho K \to \pi K^\ast$, 
$\pi \phi \to K K^\ast$ (or $K^\ast K$),  
$\pi \phi \to K^\ast K^\ast$, $\rho \phi \to K K^\ast$ (or $K^\ast K$) and  
$\rho \phi \to K^\ast K^\ast$. Our calculations give the 
$\sqrt s$-dependent cross sections for the $\phi$ absorption reactions 
including the notable $I=1~\rho \phi \to K^\ast K^\ast$ and the average cross 
sections that depend on temperature of hadronic matter. 
Every endothermic reaction in Subsections 4.2, 4.3 and 4.4 
has a maximum cross section larger than the 
cross section for its corresponding inverse exothermic reaction at the same
center-of-mass energy except the reaction of $I=3/2~\pi K^\ast \to \rho K$.
Subsequently, $K^\ast$ and $\rho$ from the
creation reactions with the appreciable cross sections are 
expected to compensate for the loss due to decay in hadronic 
matter and to make $K^\ast$ and $\rho$ observable in Au+Au collisions.

\vspace{0.5cm}
\leftline{\bf 5. Summary}
\vspace{0.5cm}

We have studied meson-meson inelastic scatterings with quark-antiquark
relative-motion wave functions that are solutions of the Schr\"{o}dinger
equation with the Buchm\"{u}ller-Tye potential. The inelastic scatterings 
are governed by the quark-interchange mechanism that gives rise to the prior 
form and the post form. The post-prior discrepancy is completely negligible in
the S-wave elastic phase shifts for $I=2~\pi \pi $ and $I=3/2~K \pi $ 
scattering in spite of that the physical meson masses and 
flavor-symmetry breaking are employed throughout this work. The two forms do 
produce different cross sections of which the average is taken as the 
unpolarized cross section. It is found that the reaction of 
 $I=3/2$ $\pi K^\ast \to \rho K$ is most important among the 
endothermic nonresonant reactions. 

If the difference between $\sqrt s$ and threshold energy is the same for the
reactions of $I=2$ $\pi \pi \leftrightarrow \rho \rho$, $I=2$ $\pi \rho 
\leftrightarrow \rho \rho$, $I=1$ $K K \leftrightarrow K^\ast K^\ast$ and 
$I=1$ $K K^\ast \leftrightarrow K^\ast K^\ast$, at this $\sqrt s$ the cross 
sections for $I=1$ $K K \leftrightarrow K^\ast K^\ast$ ($I=1$ $K K^\ast 
\leftrightarrow K^\ast K^\ast$) are larger than the cross sections for 
$I=2$ $\pi \pi \leftrightarrow \rho \rho$ ($I=2$ $\pi \rho 
\leftrightarrow \rho \rho$). This means that mesonic interactions in matter
consisting of only $K$ and $K^\ast$ can be stronger than mesonic interactions
in matter consisting of only $\pi$ and $\rho$.

\vspace{0.5cm}
\leftline{\bf Acknowledgements}
\vspace{0.5cm}
We thank Y.-G. Ma and L.-W. Chen for helpful discussions. After the manuscript
was accepted for publication, we got from B.S. Zou an interesting study of
$\pi \pi$ elastic scattering which was published in Nucl. Phys. A735 (2004) 
111. This work was
supported by National Natural Science Foundation of China under Grant No.
10675079.

\newpage

\begin{figure}[htbp]
\centering
\includegraphics[scale=0.7]{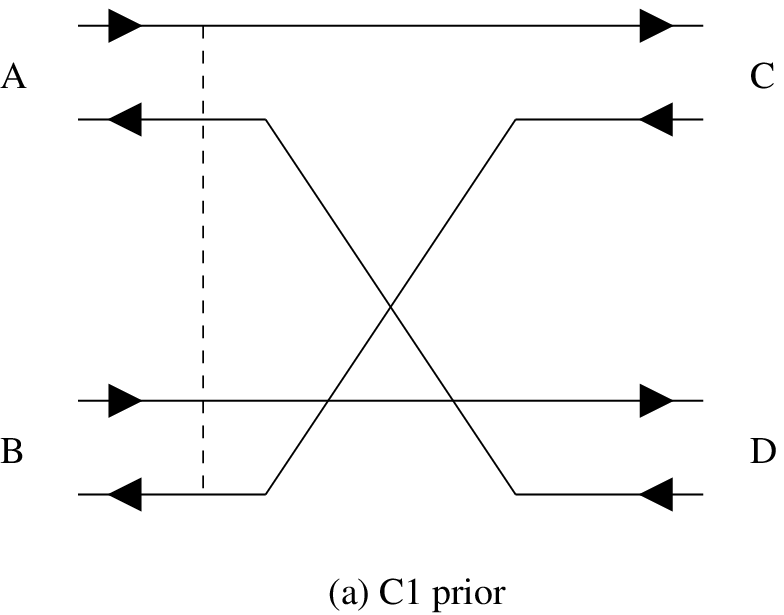}
\hspace{1.5cm}
\includegraphics[scale=0.7]{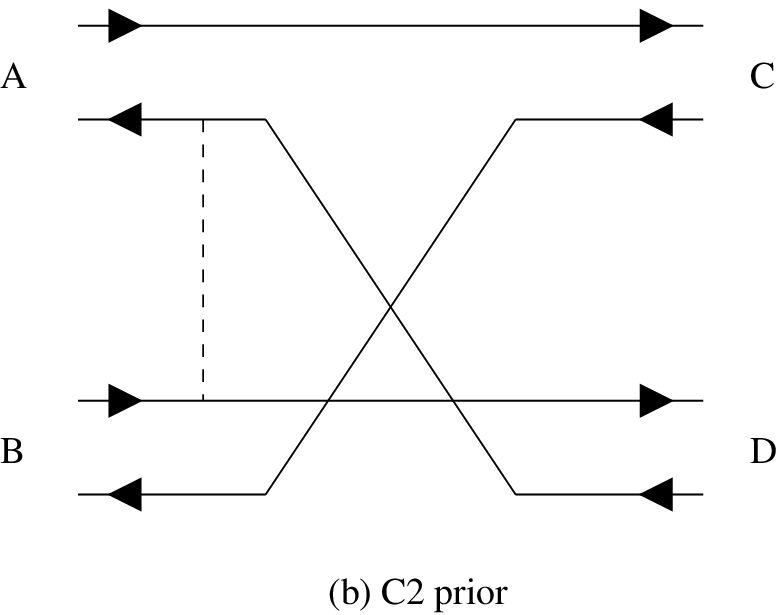}\\
\includegraphics[scale=0.7]{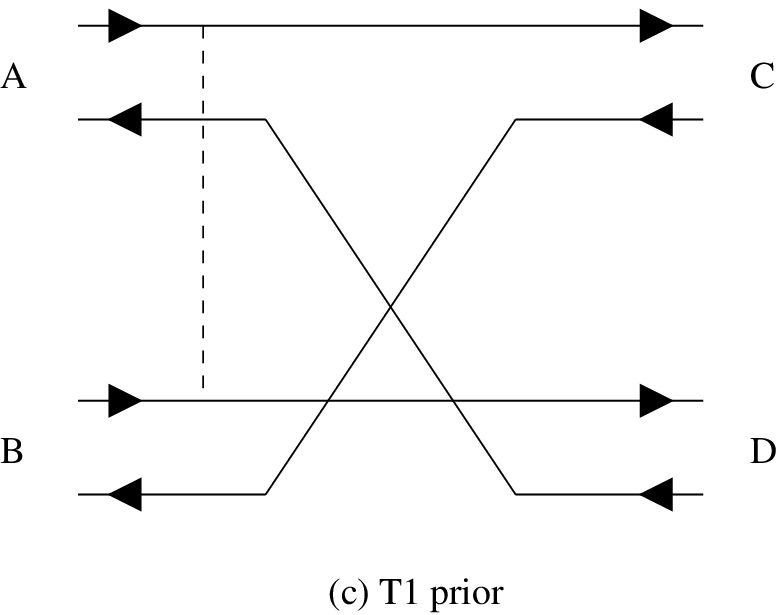}
\hspace{1.5cm}
\includegraphics[scale=0.7]{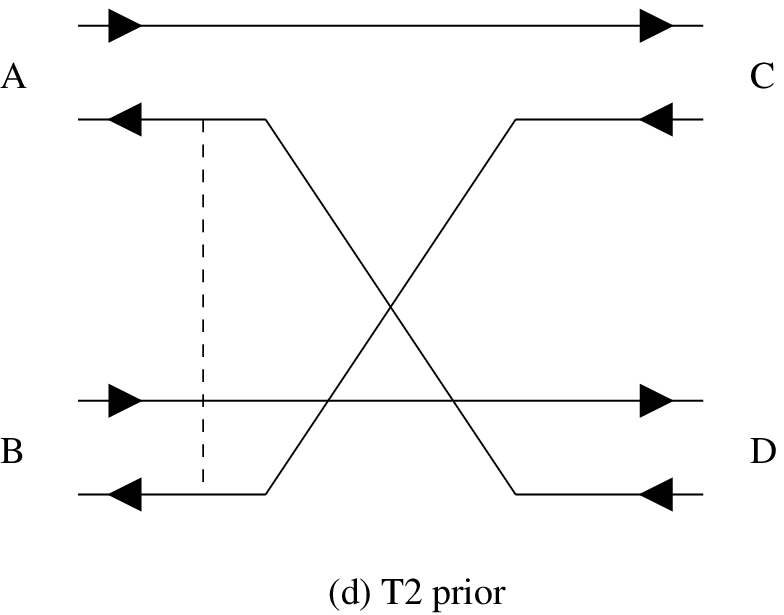}
\caption{``Prior" diagrams for the reaction $A+B\longrightarrow C+D$. Solid 
(dashed) lines represent quarks or antiquarks (gluons).}
\label{fig1}
\vspace{1.0cm}
\includegraphics[scale=0.7]{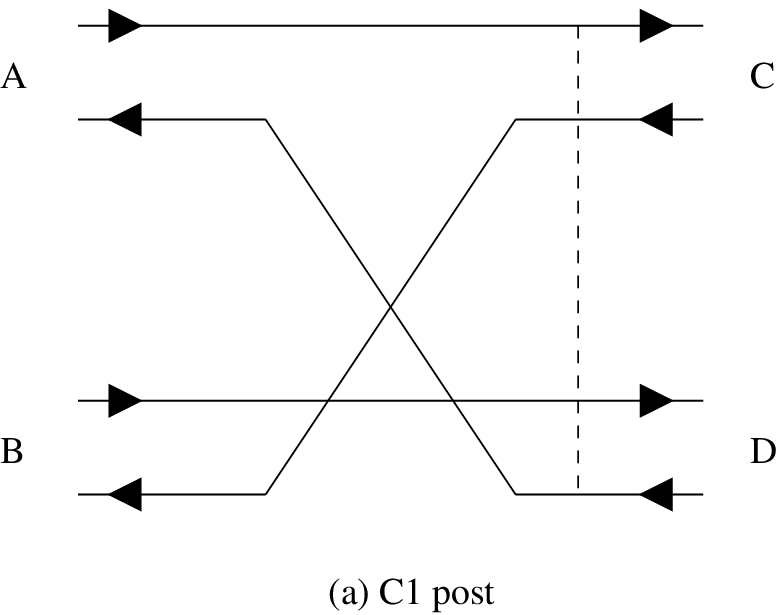}
\hspace{1.5cm}
\includegraphics[scale=0.7]{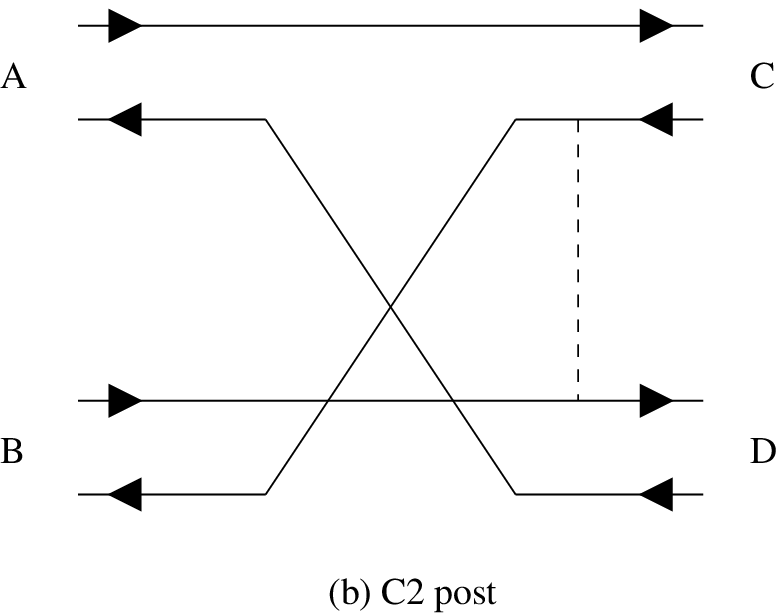}\\
\includegraphics[scale=0.7]{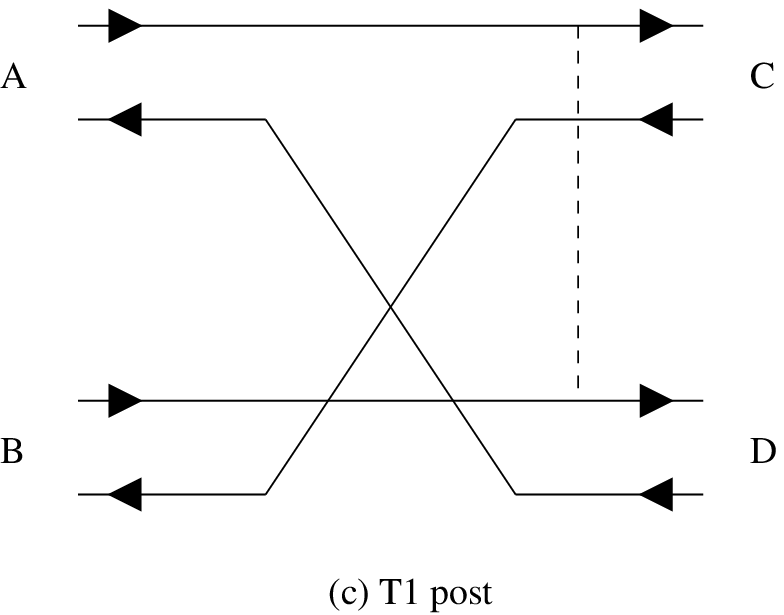}
\hspace{1.5cm}
\includegraphics[scale=0.7]{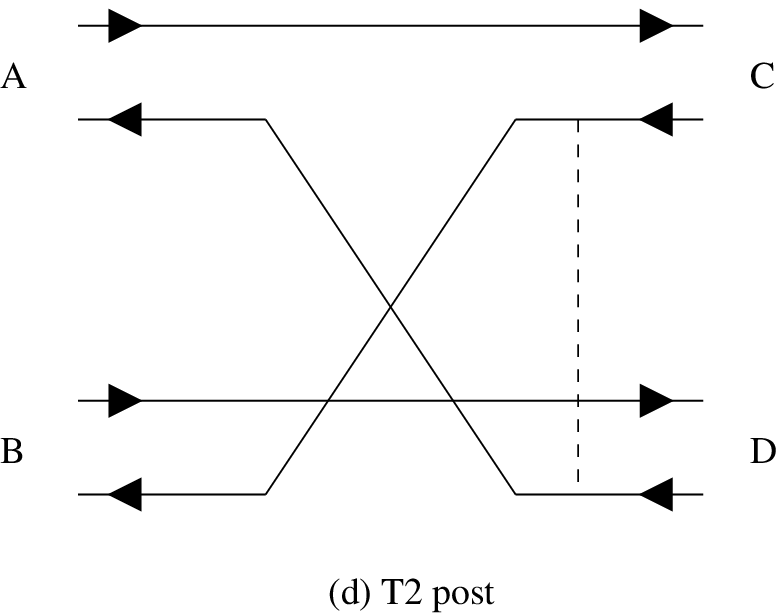}
\caption{``Post" diagrams. Solid (dashed) lines represent quarks or antiquarks 
(gluons).}
\label{fig2}
\end{figure}

\begin{figure}[htbp]
\centering
\includegraphics[width=9cm]{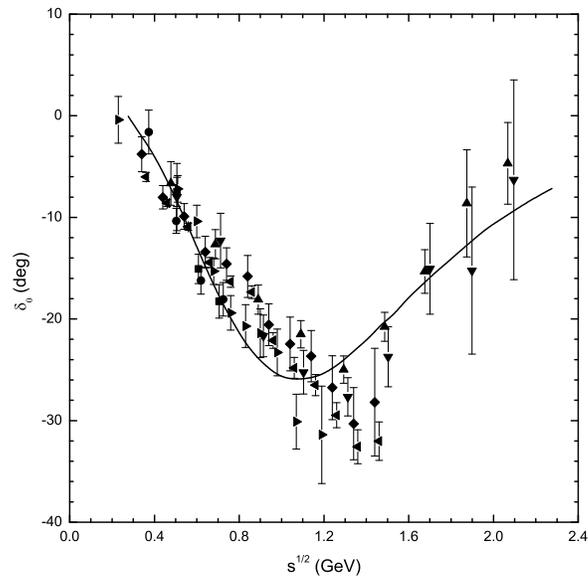}%
\caption{Theoretical S-wave $I=2~\pi\pi$ phase shift (solid curve)
compared to experimental data. Squares: Table I of Ref. \cite{Col71};
circles: Table II of Ref. \cite{Col71}; triangles up: OPE data of Ref. 
\cite{Dur73}; triangles down: OPE-DP data of Ref. \cite{Dur73}; diamonds: data
of set A of Ref. \cite{Hoo77}; triangles left: data of set B of Ref. 
\cite{Hoo77}; triangles right: Ref. \cite{Los74}. }
\label{fig3}
\end{figure}

\begin{figure}[htbp]
\centering
\includegraphics[width=10cm]{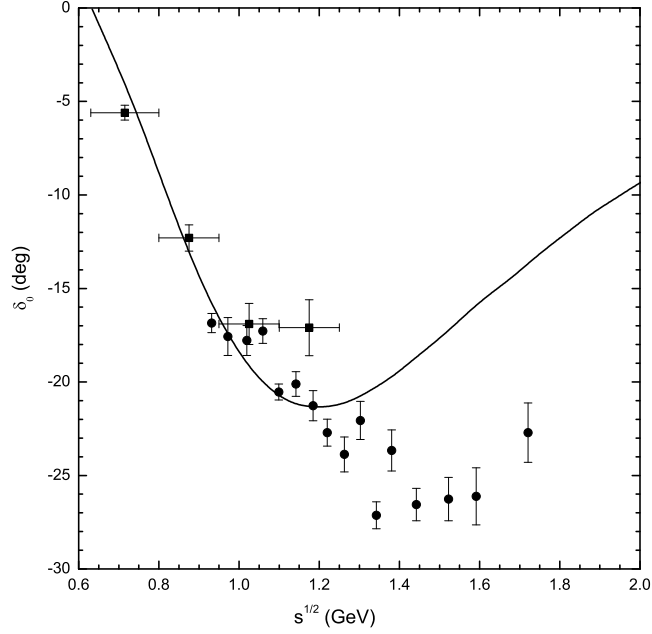}
\caption{Theoretical S-wave $I=3/2~K\pi$ phase shift (solid curve) compared 
to experimental data. Squares: Ref. \cite{Jon73}; circles: Ref. \cite{Est78}.}
\label{fig4}
\end{figure}

\begin{figure}[htbp]
\centering
\includegraphics[scale=1.0]{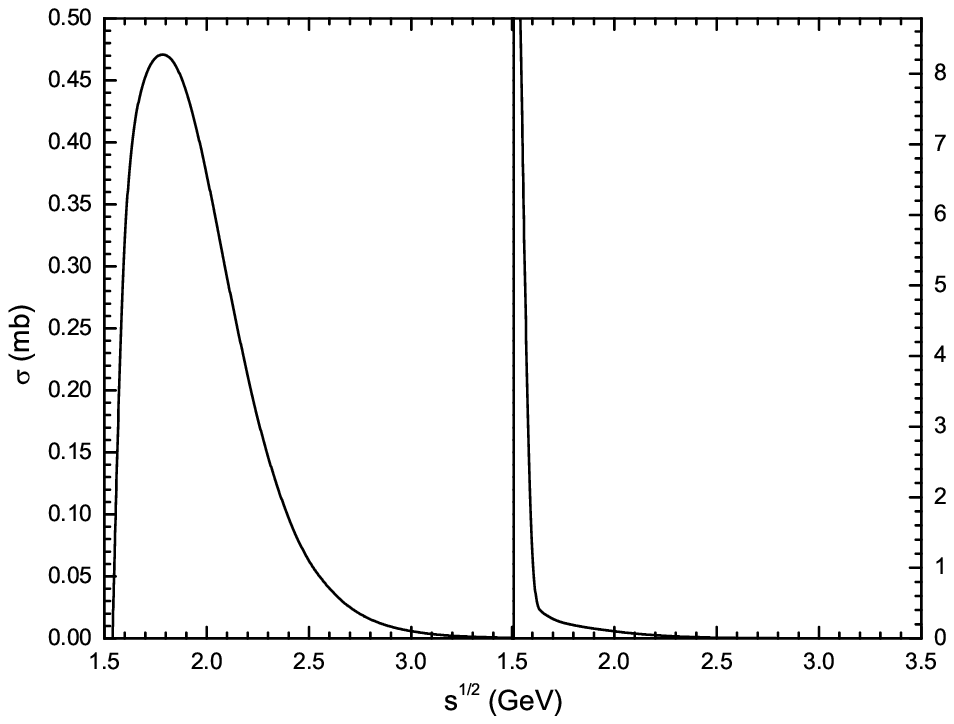}%
\caption{Cross sections for the $I=2~\pi \pi \to \rho \rho $ in the
left panel and for the $I=2~\rho \rho \to \pi \pi$ in the right panel.}
\label{fig5}
\end{figure}

\begin{figure}[htbp]
\centering
\includegraphics[scale=1.0]{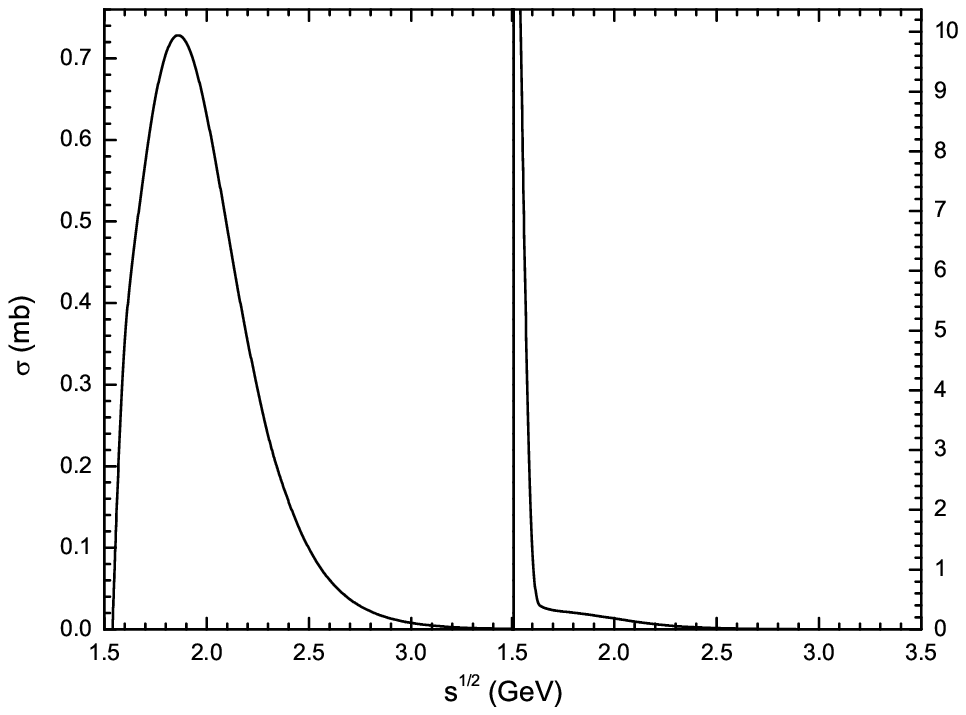}
\caption{Cross sections for the $I=2~\pi \rho \to \rho \rho $ in the
left panel and for the $I=2~\rho \rho \to \pi \rho $ in the right panel.}
\label{fig6}
\end{figure}

\begin{figure}[htbp]
\centering
\includegraphics[scale=1.0]{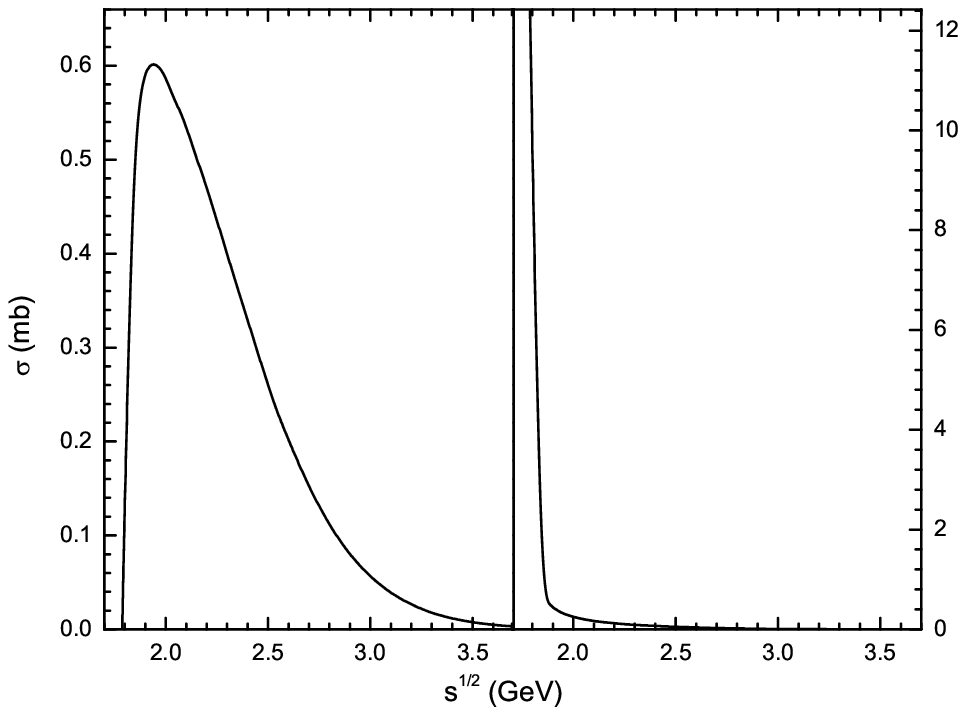}%
\caption{Cross sections for the $I=1~K K \to K^\ast K^\ast $ in the
left panel and for the $I=1~K^\ast K^\ast \to K K $ in the right panel.}
\label{fig7}
\end{figure}

\begin{figure}[htbp]
\centering
\includegraphics[scale=1.0]{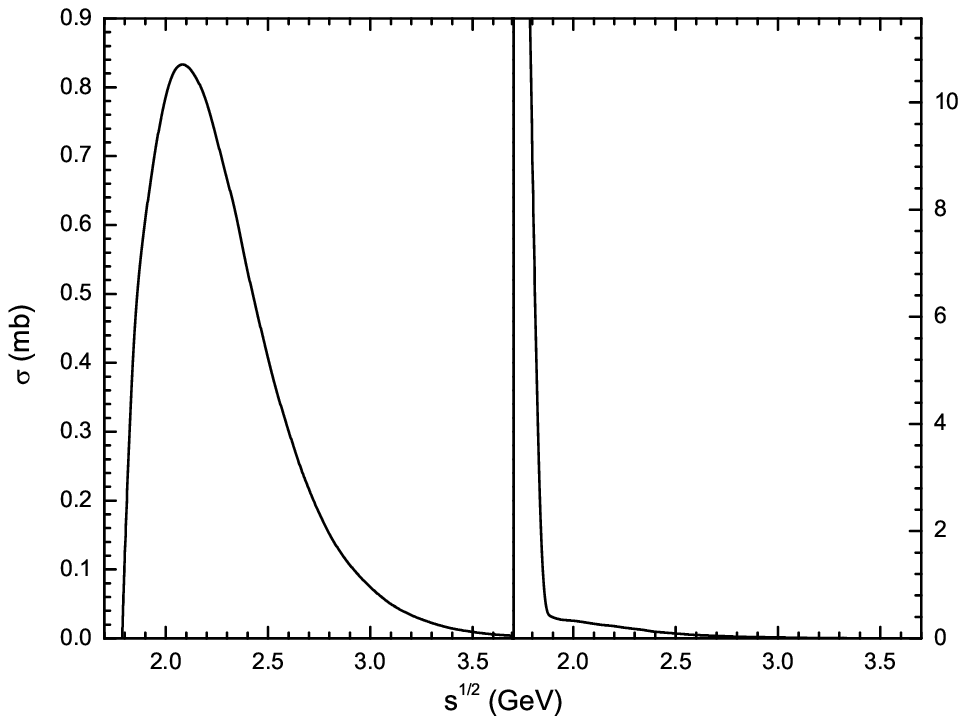}
\caption{Cross sections for the $I=1~K K^\ast \to K^\ast K^\ast $ in the left 
panel and for the $I=1~K^\ast K^\ast \to K K^\ast $ in the right panel.}
\label{fig8}
\end{figure}

\begin{figure}[htbp]
\centering
\includegraphics[scale=1.0]{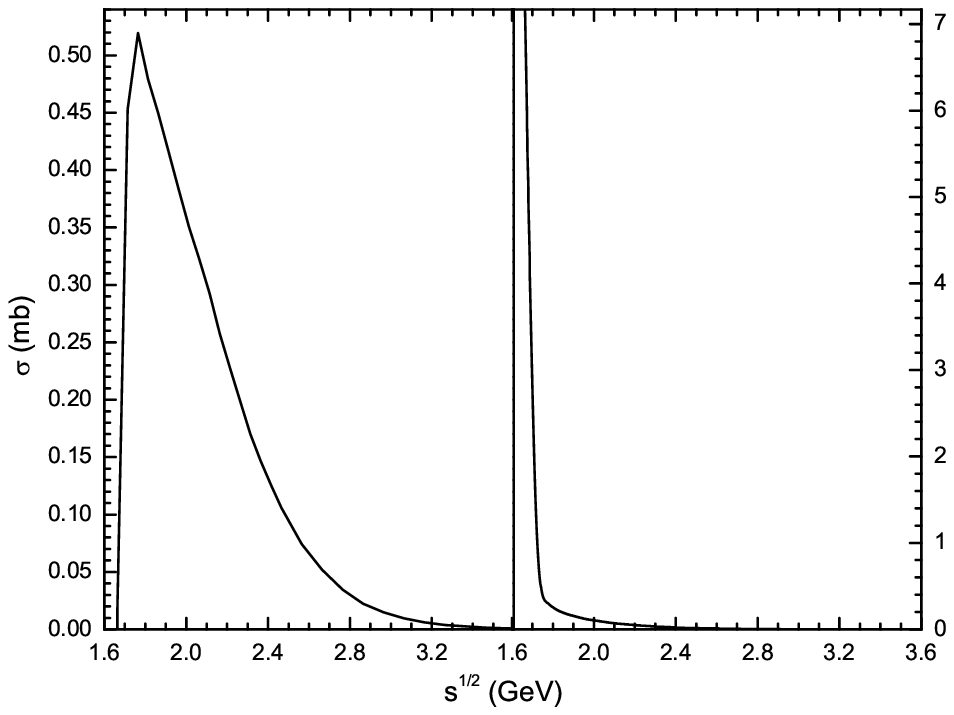}%
\caption{Cross sections for the $I=3/2~\pi K \to \rho K^\ast$ in the
left panel and for the $I=3/2~\rho K^\ast \to \pi K$ in the right panel.}
\label{fig9}
\end{figure}

\begin{figure}[htbp]
\centering
\includegraphics[scale=1.0]{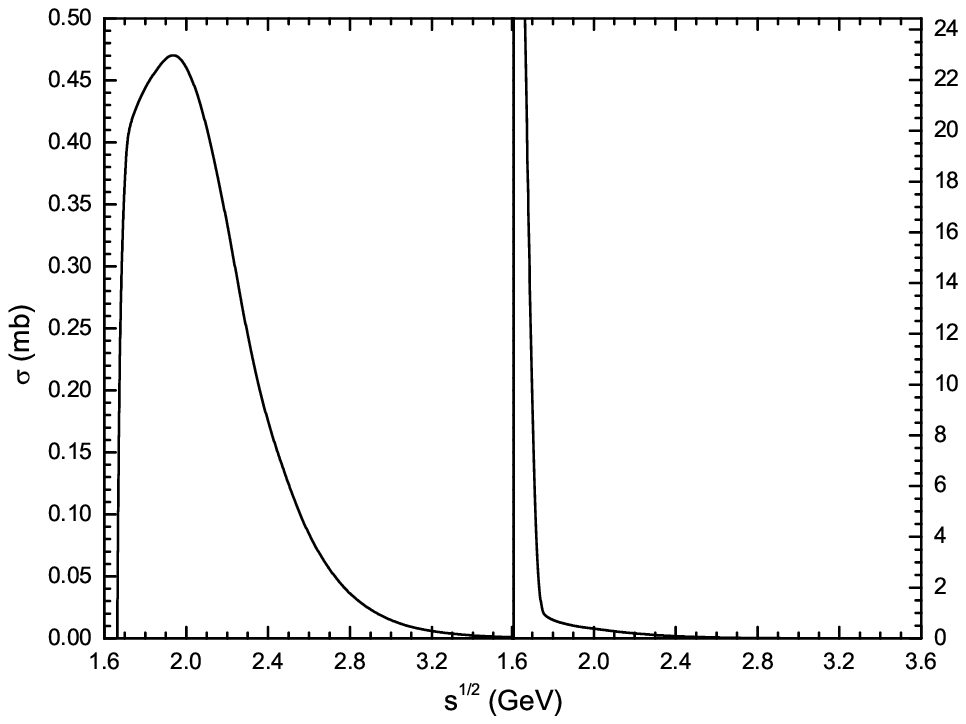}
\caption{Cross sections for the $I=3/2~\pi K^\ast \to \rho K^\ast$
in the left panel and for the  $I=3/2~\rho K^\ast \to \pi K^\ast $ in the
right panel.}
\label{fig10}
\end{figure}

\begin{figure}[htbp]
\centering
\includegraphics[scale=1.0]{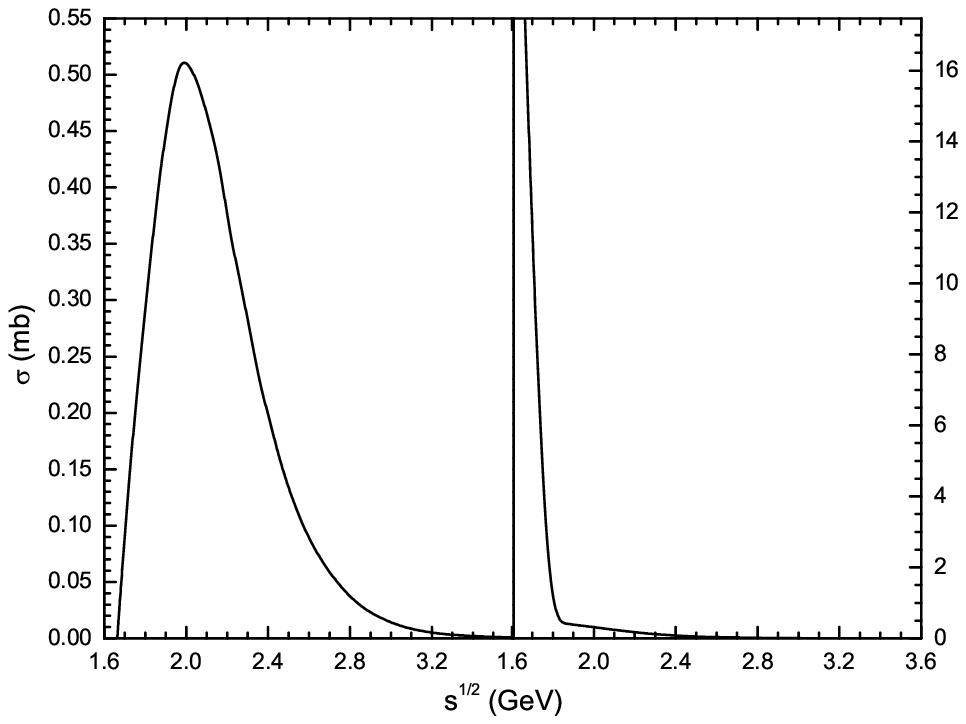}
\caption{Cross sections for the $I=3/2~\rho K \to \rho K^\ast $ in
the left panel and for the $I=3/2~\rho K^\ast \to \rho K $ in the right panel.}
\label{fig11}
\end{figure}

\begin{figure}[htbp]
\centering
\includegraphics[scale=1.0]{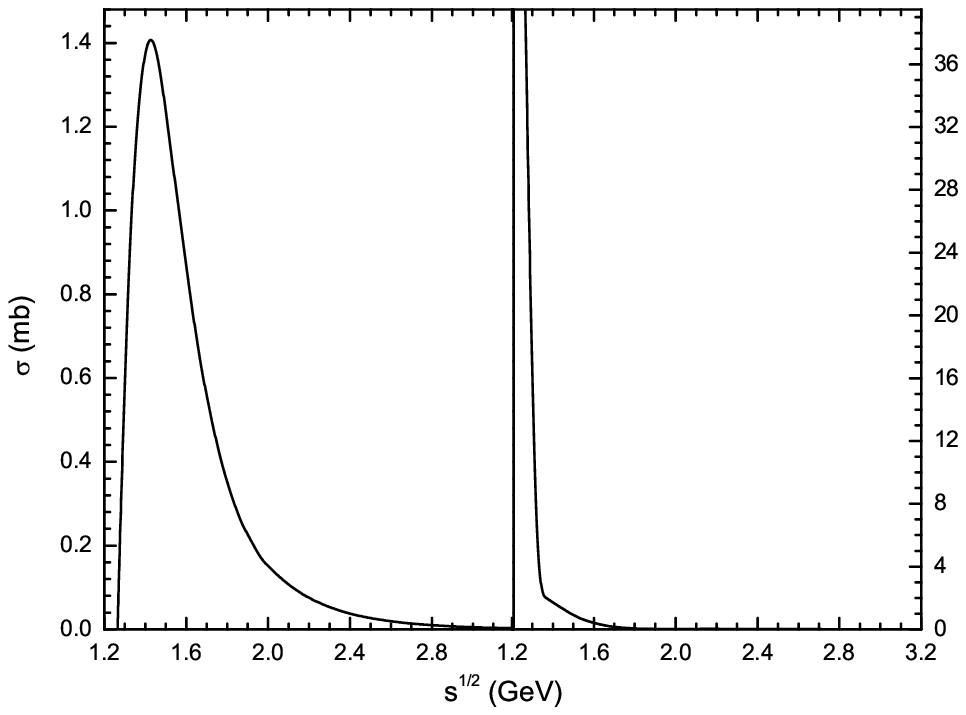}%
\caption{Cross sections for the $I=3/2~\pi K^\ast \to \rho K$ in the
left panel and for the $I=3/2~\rho K \to \pi K^\ast $ in the right panel.}
\label{fig12}
\end{figure}

\begin{figure}[htbp]
\centering
\includegraphics[scale=1.0]{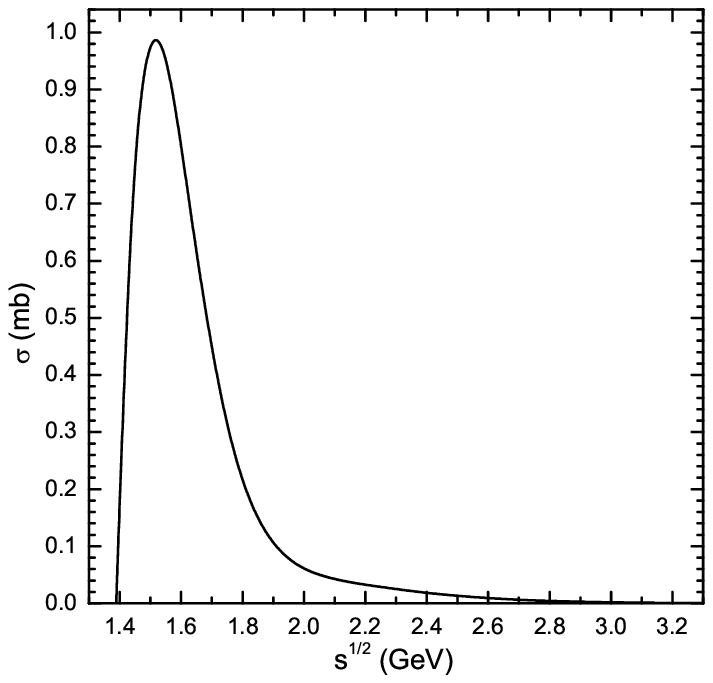}
\caption{Cross section for the $I=1~\pi \phi \to K K^\ast$ (or $K^\ast K$).}
\label{fig13}
\end{figure}

\begin{figure}[htbp]
\centering
\includegraphics[scale=1.0]{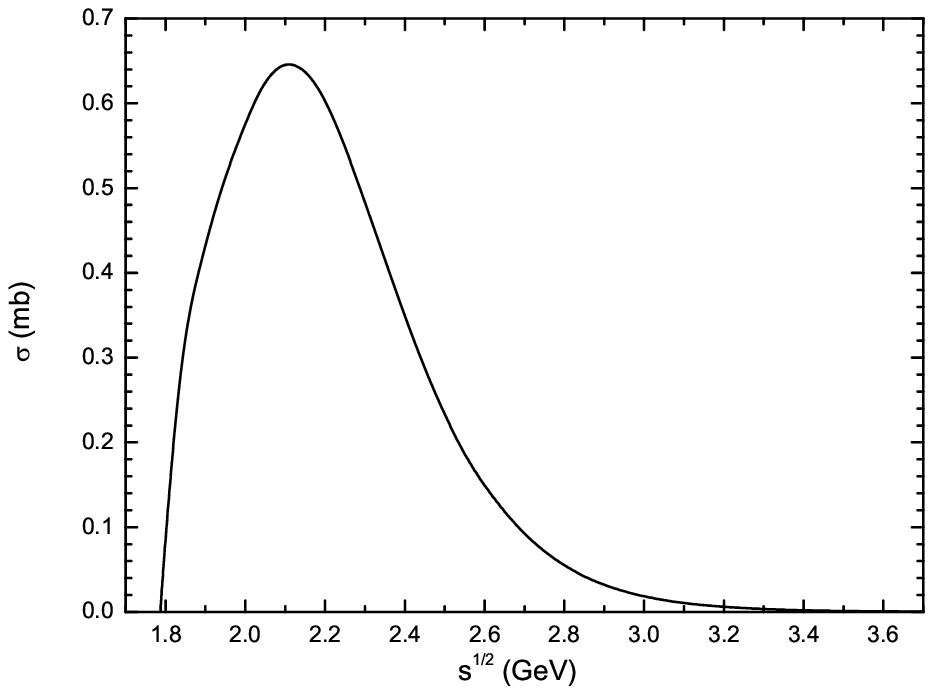}
\caption{Cross section for the $I=1~\pi \phi \to K^\ast K^\ast$.}
\label{fig14}
\end{figure}

\begin{figure}[htbp]
\centering
\includegraphics[scale=1.0]{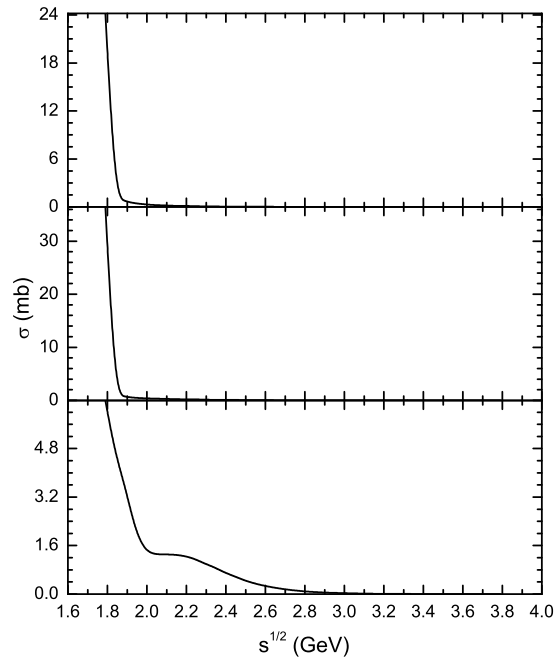}%
\caption{Solid line in the upper (middle, lower) panel shows cross
section for the $I=1~\rho \phi \to K K$ ($I=1~\rho \phi \to K
K^\ast$ (or $K^\ast K$), $I=1~\rho \phi \to K^\ast K^\ast$).}
\label{fig15}
\end{figure}

\end{document}